\documentclass[aps,10pt,prd,superscriptaddress,preprintnumbers,%
showkeys,nofootinbib,showpacs,twocolumn]{revtex4-1}
\usepackage{slashed}
\usepackage{eucal} 
\usepackage{bm}
\usepackage{graphicx}
\usepackage{amsmath,amssymb,amsfonts,latexsym}

\usepackage[hypertexnames=false]{hyperref}
\begin{document}
\author{June-Young Kim}
\email[ E-mail: ]{jykim@jlab.org}
\affiliation{Theory Center, Jefferson Lab, Newport News, VA 23606, USA}

\title{Chiral-odd generalized parton distributions in the large-$N_{c}$ limit of QCD: Next-to-leading-order contributions}
\begin{abstract}
We investigate the nucleon’s chiral-odd generalized parton distribution functions (GPDs) in the large-$N_c$ limit of QCD. Extending previous work~\cite{Kim:2024ibz} on the leading-order contribution in the $1/N_c$ expansion, we focus on the next-to-leading-order contributions and provide a complete set of flavor-singlet and flavor-non-singlet chiral-odd GPDs. This study includes the derivation of the spin-flavor structure of the baryon matrix element of the chiral-odd operator, the proof of the polynomiality property and associated sum rules, and numerical estimates based on the gradient expansion. The spin-flavor structure of the nucleon matrix element is interpreted through a multipole expansion in the transverse momentum transfer, leading to the definition of multipole mean-field GPDs. Using these GPDs as the basis of our analysis, we take their $m$-th moments and demonstrate the polynomiality property within the mean-field framework, making use of discrete symmetries in the proof. In particular, the first moments ($m = 1$) of the GPDs are related to the nucleon tensor form factors, as generally required. By performing a gradient expansion, we compute the chiral-odd GPDs and present numerical estimates. Our results show good agreement with lattice QCD predictions.
\end{abstract}
\maketitle
\tableofcontents
%
%

\section{Introduction}

Generalized parton distributions (GPDs) are a powerful tool for investigating the internal structure of the nucleon, unifying the concepts of parton distribution functions and nucleon form factors (see Refs.~\cite{Goeke:2001tz,Diehl:2003ny,Belitsky:2005qn,Boffi:2007yc} for reviews). In particular, chiral-odd GPDs contain information about the distribution of transversely polarized partons within the nucleon. In the transverse coordinate representation, they offer a spatial interpretation of how such partons are distributed in impact parameter space~\cite{Burkardt:2005hp,Diehl:2005jf}. Despite their importance, chiral-odd GPDs remain less well understood than their chiral-even counterparts, due to their suppression in leading-twist hard exclusive processes and the relatively limited theoretical understanding.

Although chiral-odd GPDs enter at subleading twist, they can be accessed experimentally via the twist-3 mechanism in exclusive pseudoscalar meson production, through their coupling to chiral-odd meson distribution amplitudes~\cite{Ahmad:2008hp, Goloskokov:2009ia, Goloskokov:2011rd, Goldstein:2012az}. This mechanism, though formally suppressed, yields sizable contributions at intermediate momentum transfers, consistent with JLab 6 GeV data~\cite{CLAS:2012cna,CLAS:2014jpc,CLAS:2017jjr}. These findings have motivated further studies at JLab 12 GeV~\cite{CLAS:2023wda}, CERN COMPASS~\cite{COMPASS:2019fea}, and the future Electron-Ion Collider~\cite{AbdulKhalek:2021gbh}, highlighting the need for precise theoretical input to support current and forthcoming experiments.

Spontaneous chiral symmetry breaking (S$\chi$SB) is a fundamental phenomenon underlying the nonperturbative structure of the nucleon. One of the most plausible mechanisms responsible for S$\chi$SB is the QCD instanton vacuum. At a low normalization scale, the QCD vacuum is posited to be populated by an ensemble of instantons, with an average instanton size of $\bar{\rho} \approx 1/3~\mathrm{fm}$ and an average inter-instanton distance of $\bar{R} \approx 1~\mathrm{fm}$. These instantons induce localized zero modes of fermion fields with definite chirality. The delocalization of these modes leads to the spontaneous breaking of chiral symmetry, dynamically generating a fermion mass (see Refs.~\cite{Diakonov:2002fq, Schafer:1996wv} for a review). Upon bosonization of the fermionic degrees of freedom, the Goldstone boson—identified with the massless pion—emerges as an effective degree of freedom. Consequently, an effective chiral action capturing the essential nonperturbative dynamics of QCD emergies~\cite{Diakonov:1985eg, Diakonov:1995qy, Kacir:1996qn}.

By employing the semiclassical approximation to the effective chiral action—justified in the large-$N_c$ limit of QCD—the internal dynamics of the nucleon can be systematically studied. In this limit, quantum fluctuations of the pion field are suppressed, and the pion field is governed by the solution to the classical equations of motion. This framework is commonly referred to as the mean-field picture. Within this framework, the nucleon is described as a bound state of $N_c$ valence quarks interacting with a self-consistent pion mean field~\cite{Witten:1979kh} (see also Refs.~\cite{Diakonov:1987ty, Wakamatsu:1990udFdia, Christov:1995vm} for a review). In particular, a key feature of the $1/N_c$ expansion is the emergence of spin-flavor symmetry~\cite{Gervais:1983wq,Dashen:1993jt}, which is naturally realized in the mean-field picture through the so-called hedgehog symmetry~\cite{Pauli:1942kwa, Adkins:1983ya}—a minimal generalization of spherical symmetry that preserves simultaneous rotations in ordinary space and isospin space. While treating the mean-field solution as an abstract object, the spin-flavor structure of QCD matrix elements has been investigated in various contexts~\cite{Pobylitsa:2000tt, Goeke:2001tz, Pobylitsa:2003ty, Schweitzer:2016jmd, Panteleeva:2020ejw, Kim:2024ibz}. In addition, based on self-consistent mean-field solutions, studies have explored tensor form factors~\cite{Kim:1995bq, Kim:1996vk, Ledwig:2010zq, Ghim:2025gqo} and transversity parton distributions~\cite{Schweitzer:2001sr, Wakamatsu:2008ki}. Only very recently, a comprehensive study~\cite{Kim:2024ibz} of chiral-odd GPDs at leading order in the $1/N_c$ expansion has been conducted, including the derivation of their spin-flavor structure and proofs of polynomiality and sum rules. However, the spin-flavor structure at leading-order~(LO) is not sufficient to account for the complete set of chiral-odd GPDs in both the flavor-singlet and flavor-non-singlet sectors.

In this work, we extend the analysis of the chiral-odd GPDs to next-to-leading order~(NLO) in the $1/N_c$ expansion, completing the spin-flavor structures that were missing at LO in both the flavor-singlet and flavor-non-singlet sectors. This study also includes the proof of the polynomiality property and the corresponding sum rules in the single-particle representation. In addition, we provide numerical predictions for the chiral-odd GPDs using the gradient expansion. 

The non-perturbative aspects of the chiral-odd GPDs have also been explored using various quark model approaches, including light-front quark models~\cite{Pasquini:2005dk, Chakrabarti:2008mw, Lorce:2011dv, Chakrabarti:2015ama}, the bag model~\cite{Tezgin:2024tfh}, and the basis light-front quantization model~\cite{Kaur:2023lun, Liu:2024umn}. The tensor form factors have been calculated within the framework of light-cone QCD sum rules~\cite{Aliev:2011ku}. In lattice QCD, the moments of GPDs have been computed in Refs.~\cite{QCDSF:2006tkx, Park:2021ypf, Alexandrou:2022dtc}, while the $x$-dependence of the chiral-odd GPDs~\cite{Alexandrou:2021bbo, Alexandrou:2021oih, Bhattacharya:2025yba} has been extracted by computing quasi-GPDs and performing perturbative matching.

This paper is organized as follows. In Sec.~\ref{sec:gf}, we define the chiral-odd GPDs and discuss their polynomiality properties, along with the sum rules that relate them to tensor form factors in a general framework. Section~\ref{sec:pmfa} introduces the pion mean-field picture. Based on this approach, we derive the spin-flavor structure of the matrix element of the chiral-odd operator. Through the multipole expansion, we then obtain the corresponding mean-field GPDs in the $1/N_{c}$ expansion. In Sec.~\ref{sec:mimfp}, we compute the $m$-th moments of the mean-field GPDs. Using discrete symmetries and these moment results, we demonstrate their polynomiality and verify the associated sum rules. In Sec.~\ref{sec:ge}, the chiral-odd GPDs are analyzed using the gradient expansion. In Sec.~\ref{num}, we present numerical estimates of the chiral-odd GPDs and compare them with lattice QCD results. The final section, Sec.~\ref{sec:sac}, summarizes our findings and draws conclusions.

\section{General formalism \label{sec:gf}}
The matrix element of the chiral-odd operator between nucleon states is defined as:
\begin{align}
& \mathcal{M}^{q}_{\mathrm{GPDs}}[i\sigma^{+j}] =
P^{+}\int \frac{d z^{-}}{2\pi} e^{i x P^{+} z^{-} } 
\nonumber \\
&\hspace{-0.1cm} \times \langle p', s' | \bar{\psi} \left(-\frac{z}{2}\right) 
\left[-\frac{z}{2}, \frac{z}{2} \right] i\sigma^{+j} \psi \left(\frac{z}{2}\right) |  p, s \rangle 
\bigg{|}_{z^{+}, \bm{z}_\perp = 0}.
\label{eq:General_ME}
\end{align}
$\psi$ and $\bar{\psi}$ denote the quark field operators, $\sigma^{\mu\nu} \equiv (i/2) [\gamma^\mu, \gamma^\nu]$, and the components of a light-cone four-vector are defined as $v^{\pm} \equiv (v^{0} \pm v^{3})/\sqrt{2}$ and $\bm{v}_{\perp} = (v^{1}, v^{2})$. The space-time separation $z$ between the quark fields is light-like, satisfying $z^{2} = 0$, and $[-z/2, z/2]$ represents the gauge link that ensures gauge invariance. The initial and final nucleon four-momenta are $p$ and $p'$, with their average and difference defined as $P \equiv (p' + p)/2$ and $\Delta \equiv p' - p$, respectively. The variables $s$ and $s'$ denote the spin polarizations of the initial and final nucleon states.

The matrix element of Eq.~\eqref{eq:General_ME} is parameterized as
\begin{align}
& \mathcal{M}_{\mathrm{GPDs}}[i\sigma^{+j}] = 
\bar{u}' \left[ i \sigma^{+j} \, H_{T}
+ \frac{P^{+} \Delta^{j} - \Delta^{+} P^{j}}{M^{2}_{N}} \, \tilde{H}_{T} \right.
\nonumber \\
& \left. + \frac{\gamma^{+}\Delta^{j} -\Delta^{+} \gamma^{j} }{2M_{N}} \, E_{T}
+ \frac{\gamma^{+} P^{j} - P^{+} \gamma^{j}}{M_{N}} \, \tilde{E}_{T} 
\right] u.
\label{eq:General_ME_GPDs}
\end{align}
$u \equiv u(p, s)$ and $\bar{u}' \equiv \bar{u}(p', s')$ are the Dirac spinors corresponding to the initial and final nucleon states, respectively. These spinors are normalized such that $\bar{u}u = \bar{u}'u' = 2M_{N}$, where $M_{N}$ denotes the nucleon mass. The functions $H_{T}$, $\tilde{H}_{T}$, $E_{T}$, and $\tilde{E}_{T}$ are chiral-odd GPDs. They depend on the longitudinal momentum fraction $x$ carried by partons, the skewness $\xi \equiv -\Delta^{+}/2P^{+}$, and the squared four-momentum transfer $t \equiv \Delta^{2}$. These GPDs also depend on the renormalization scale of the QCD operator, although this dependence is not shown explicitly. Moreover, both the QCD operator in Eq.~\eqref{eq:General_ME} and the chiral-odd GPDs in Eq.~\eqref{eq:General_ME_GPDs} depend on the quark flavor component $f = (u, d)$. In what follows, we will specify the flavor index only when necessary.

The time-reversal symmetry of the matrix element~\eqref{eq:General_ME} imposes a constraint on the behavior of the chiral-odd GPDs under the transformation $\xi \to -\xi$:
\begin{subequations}
\label{xi_parity}
\begin{align}
&F(x, \xi, t) = +F(x, -\xi, t) \quad \mathrm{for} \quad F=H_{T},\tilde{H}_{T},E_{T},
\\
&F(x, \xi, t) = -F(x, -\xi, t) \quad \mathrm{for} \quad F=\tilde{E}_{T}.
\end{align}
\end{subequations}
Most importantly, the partonic operator in Eq.~\eqref{eq:General_ME} can be expanded in powers of the light-like separation, generating a tower of local operators with increasing spin. Integrating the $x^{m-1}$-weighted partonic operator over $x$ yields a local operator of spin $m$, whose matrix elements are parameterized by generalized form factors. Consequently, the $m$-th moments of the GPDs can be expressed as polynomials in $\xi$, with coefficients given by generalized form factors that depend on $t$ \cite{Hagler:2004yt}:
\begin{subequations}
\label{eq:polynomiality}
\begin{align}
&\int^{1}_{-1} dx \, x^{m-1} H_{T}(x,\xi,t) 
= \sum^{m-1}_{\substack{i=0 \\ \mathrm{even}}} (-2\xi)^{i} A_{T m,i}(t), 
\\
&\int^{1}_{-1} dx \, x^{m-1} \tilde{H}_{T}(x,\xi,t) 
= \sum^{m-1}_{\substack{i=0 \\ \mathrm{even}}} (-2\xi)^{i} \tilde{A}_{T m,i}(t),
\\
&\int^{1}_{-1} dx \, x^{m-1} E_{T}(x,\xi,t) 
= \sum^{m-1}_{\substack{i=0 \\ \mathrm{even}}} (-2\xi)^{i} B_{T m,i}(t), 
\label{polynomiality_E_T}
\\ 
&\int^{1}_{-1} dx \, x^{m-1} \tilde{E}_{T}(x,\xi,t) 
= \sum^{m-1}_{\substack{i=0 \\ \mathrm{odd}}} (-2\xi)^{i} \tilde{B}_{T m,i}(t).
\end{align}
\end{subequations}
The polynomials are either even or odd functions of $\xi$, as dictated by Eq.~\eqref{xi_parity}. The degree of the polynomial representing the $m$-th moment is at most $m - 1$, with the precise degree depending on the parity of $m - 1$ and the specific GPD under consideration. The coefficients of these polynomials depend solely on $t$ and are referred to as generalized tensor form factors. The polynomiality property given in Eq.~\eqref{eq:polynomiality} plays a fundamental role in determining the structure of the GPDs.

The first moments of the chiral-odd GPDs are related to the nucleon tensor form factors. The nucleon matrix element of the local tensor operator is parametrized as
\begin{align}
& \mathcal{M}_{\mathrm{FFs}}[i\sigma^{\mu \nu}]
= \langle p', s' | \bar{\psi}(0) i\sigma^{\mu \nu} \psi (0) | p, s \rangle
\nonumber \\[1ex]
&= \bar{u}' \left[ i \sigma^{\mu \nu} \, H_{T}(t)
+ \frac{P^{\mu} \Delta^{\nu} - \Delta^{\mu} P^{\nu}}{M^{2}_{N}} \, \tilde{H}_{T}(t)
\right.
\nonumber \\
& \left. + \; \frac{\gamma^{\mu}\Delta^{\nu} -\Delta^{\mu} \gamma^{\nu} }{2M_{N}} \, E_{T}(t) \right] u.
\label{eq:General_ME_tensor}
\end{align}
By comparing Eq.~\eqref{eq:General_ME} and Eq.~\eqref{eq:General_ME_tensor}, one obtains the relations
\begin{subequations}
\label{eq:first_Mel}
\begin{align}
&\int^{1}_{-1} dx H_{T}(x, \xi, t) = A_{T10}(t) 
\equiv H_{T}(t), 
\\
&\int^{1}_{-1} dx \tilde{H}_{T}(x, \xi, t) = \tilde{A}_{T10}(t) \equiv \tilde{H}_{T}(t),
\\
&\int^{1}_{-1} dx E_{T}(x, \xi, t) = B_{T10}(t) \equiv 
E_{T}(t), 
\\
&\int^{1}_{-1} dx \tilde{E}_{T}(x, \xi, t)  = 0.
\label{E_T_tilde_first_moment}
\end{align}
\end{subequations}

In the forward limit, where $\xi \rightarrow 0$ and $|t| \rightarrow 0$, the chiral-odd GPD $H_T$ reduces to the transversity parton distribution function:
\begin{align}
H_{T} (x, \xi=0, t=0) = h_{1} (x).
\end{align}
Its first moment corresponds to the nucleon’s tensor charge:
\begin{align}
& \int^{1}_{-1} dx \, H_{T} (x, \xi=0, t=0) 
\nonumber \\
&= \int^{1}_{-1} dx \, h_{1} (x) =  H_{T}(0) = g_{T}.
\end{align}
In the context of exclusive pseudoscalar meson production, it is convenient to consider the linear combination of chiral-odd GPDs defined as
\begin{align}
\bar{E}_{T} \equiv E_{T}+2\tilde{H}_{T}.
\end{align}
The first moment of $\bar{E}_T$ in the forward limit yields the nucleon’s anomalous tensor magnetic moment:
\begin{align}
& \int^{1}_{-1} dx \, \bar{E}_{T} (x, \xi=0, t=0) 
\nonumber \\
& = E_{T}(0)+2\tilde{H}_{T}(0) = \kappa_{T}.
\end{align}
These results show that chiral-odd GPDs encode information about fundamental nucleon properties as probed by local tensor operators. Further discussions on the second moments of chiral-odd GPDs and their mechanical interpretations can be found in Refs.~\cite{Burkardt:2005hp, Burkardt:2006ev, Bhoonah:2017olu}.

\section{Pion mean-field approach  \label{sec:pmfa}}

\subsection{Effective chiral theory}
Nonperturbative dynamics of baryon structure can be studied using a mean-field approach based on an effective chiral theory. This effective theory emerges from the QCD instanton vacuum (see Refs.~\cite{Diakonov:1987ty, Diakonov:2002fq}), where the QCD vacuum is dominated by an ensemble of instantons with a dilute density proportional to $(\bar{\rho}/\bar{R})^{4} \approx 0.01$. Here, $\bar{\rho} \approx 1/3$~fm and $\bar{R} \approx 1$~fm denote the average instanton size and the average interdistance between instantons, respectively. Within this instanton ensemble, light quarks interact with the instantons and acquire a momentum-dependent dynamical mass $M F^{2}(p^{2})$, which drops to zero for momenta $p \gg \bar{\rho}^{-1}$. After bosonization, the pion—identified as the massless Goldstone boson—emerges as an effective degree of freedom. This framework thus explains the mechanism of chiral symmetry breaking in QCD.

At low momenta below the scale set by $\bar{\rho}^{-1}$, the dynamics of the effective theory are described by the partition function
\begin{align}
&Z_{\mathrm{eff}} = \int \mathcal{D} U \int \mathcal{D} \psi \mathcal{D} {\bar{\psi}} \exp(iS_{\mathrm{eff}}[U, \bar{\psi},\psi]), 
\end{align}
with the semi-bosonized effective action given by
\begin{align}
&S_{\mathrm{eff}}[U, \bar{\psi},\psi]  \cr
&=\int d^{4}x \, \bar{\psi}_{\alpha h} \left(i \slashed{\partial} - M F(\overleftarrow{\partial}) U^{\gamma_{5}} F(\overrightarrow{\partial}) \right)^{h}_{g}  \psi^{\alpha g},
\label{eq:par}
\end{align}
where $h, g$ and $\alpha$ denote the spin-flavor and color indices, respectively. The dynamical quark mass $M$ has a characteristic scale of $\sim \bar{R}^{-1}$ and exhibits momentum dependence governed by the quark zero-mode form factor $F$, derived from the QCD instanton vacuum at a low renormalization scale. The SU(2) bosonized chiral field is defined as
\begin{align}
U^{\gamma_{5}} &= \exp[i \bm{\pi}(x) \cdot \bm{\tau} \gamma_{5}]= \frac{1+\gamma_{5}}{2} U + \frac{1-\gamma_{5}}{2} U^{\dagger},  \\[1ex]
&\mathrm{with} \quad U= \exp[i \bm{\pi}(x) \cdot \bm{\tau} ], \nonumber
\end{align}
where $\bm{\pi}(x)$ represents the Goldstone boson field depending on spacetime coordinates, and $\bm{\tau}$ are the SU(2) flavor matrices. Here, we use the convention $\gamma_{5} \equiv i \gamma^{0} \gamma^{1} \gamma^{2} \gamma^{3}$.

\subsection{Mean-field picture}
First, we freeze the momentum dependence of the dynamical quark mass as $F(p^{2}) \approx F(0) = 1$, and regularize the effective theory by introducing an explicit cutoff $\Lambda \approx \bar{\rho}^{-1}$. Note that in Ref.~\cite{Choi:2025xha}, the mean-field picture was studied while retaining the explicit quark zero-mode form factors derived from the QCD instanton vacuum. The results of that work are generally in close agreement with the present treatment.

Next, by performing the functional integral of the semi-bosonized effective action~\eqref{eq:par} over the quark fields $\bar{\psi}$ and $\psi$, we obtain the fully bosonized effective action $S_{\mathrm{eff}}[U]$ and the Dirac operator $D(U)$:
\begin{subequations}
\label{eq:action}
\begin{align}
S_{\mathrm{eff}}[U] &=-N_{c} \, \text{Tr ln} [D(U)], \\[1ex]
D(U) &= i \partial_{0} - H(U),
\end{align}
\end{subequations}
where $\mathrm{Tr}...$ denotes a functional trace, and $H(U)$ is the one-particle Dirac Hamiltonian
\begin{align}
H(U)= -i \gamma^{0} \gamma^{k} \partial_{k} + M \gamma^{0} U^{\gamma_{5}}.
\label{hamiltonian}
\end{align}
which describes the orbital motion of single particles in the chiral field.

To proceed, we assume that the chiral field exhibits hedgehog symmetry and is static. This configuration realizes the contracted spin-flavor symmetry characteristic of baryons in the large-$N_{c}$ limit. Under hedgehog symmetry, the chiral field takes a specific spatial form in which the isospin direction $\bm{\tau}$ is aligned with the spatial direction $\bm{x}$:
\begin{align}
U(\bm{x})&= \exp[i \bm{\pi}(\bm{x}) \cdot \bm{\tau}], \nonumber \\[1ex]
\bm{\pi}(\bm{x}) &\equiv \frac{\bm{x}}{r} P(r), \quad \mathrm{with} \quad r\equiv|\bm{x}|.
\label{eq:hed}
\end{align}
The radial component of the pion field $\bm{\pi}(\bm{x})$ is characterized by the profile function $P(r)$, which satisfies the boundary conditions $P(0) = \pi$ and $P(\infty) = 0$. This field configuration preserves invariance under simultaneous spatial and isospin rotations, and thereby embodies the emergent spin-flavor symmetry of baryons in the large-$N_{c}$ limit.

The single-particle wave functions $\Phi_{n}(\bm{x})$ and their corresponding energy eigenvalues $E_{n}$ are determined by diagonalizing the Hamiltonian~\eqref{hamiltonian}:
\begin{align}
H \Phi_{n}(\bm{x})= E_{n}\Phi_{n}(\bm{x}).
\label{eq:basis}
\end{align}
Each single-particle wave function $\Phi_{n}$ is therefore characterized by its energy eigenvalue. In addition, the Hamiltonian commutes with both the grand spin operator $\bm{G} = \bm{\Sigma}/2 + \bm{T} + \bm{L}$ and the relativistic parity operator $\hat{\Pi}$:
\begin{align}
[H, \bm{G} ] = 0, \quad [H, \hat{\Pi} ] = 0,
\label{eq:basis_a}
\end{align}
where $\bm{\Sigma} = \gamma_{0} \bm{\gamma} \gamma_{5}$, $\bm{T} = \bm{\tau}/2$, and $\bm{L} = \bm{x} \times \bm{p}$ denote the spin, isospin, and orbital angular momentum operators, respectively. Given Eqs.~\eqref{eq:basis} and \eqref{eq:basis_a}, the single-particle wave function can be labeled by the corresponding quantum numbers:
\begin{align}
&|n \rangle \equiv |n = \{E_{n}, G, G_{3}, \mathcal{P}_{n} \} \rangle, \nonumber \\[1ex]
&\text{with} \quad  \Phi_{n}(\bm{x})\equiv \langle \bm{x}| n \rangle.
\label{eq:newf}
\end{align}
where $\mathcal{P}_{n} = \pm 1$ is the eigenvalue of the parity operator.

The energy spectrum includes both the upper and lower Dirac continua, which are distorted due to the presence of the chiral field. As the strength of the chiral field increases, the lowest single-particle level,
\begin{align}
&|\mathrm{lev}\rangle \equiv|n=\{E_{\mathrm{lev}},G=0,G_{3}=0, \mathcal{P}_{\mathrm{lev}}=+ \} \rangle,
\label{eq:bswf}
\end{align}
is pulled down from the upper Dirac continuum and enters the mass gap, such that $-M < E_{\mathrm{lev}} < M$. This leads to the formation of exactly one bound state level and acquires the unity baryon number $B=1$. 

The large-$N_{c}$ limit of QCD allows us to apply the mean-field approximation to determine the nucleon mass. In this framework, the energy of the system is given by the sum of the discrete level energy~\eqref{eq:bswf} and the contribution from the negative-energy continuum~\eqref{eq:newf}, with the vacuum energy subtracted:
\begin{align}
E[U] &= N_{c} E_{\mathrm{lev}} + N_{c} E_{\mathrm{cont}}, \nonumber \\[1ex]
&= N_{c} \sum_{n, \mathrm{occ}} E_{n},
\label{eq:te}
\end{align}
where “occ” denotes the occupied quark states, which include the discrete level and the negative-energy continuum. The energy of the system~\eqref{eq:te} is a functional of the chiral field configuration. The realistic/physical configuration of the chiral field is obtained by minimizing the energy functional, which corresponds to solving the saddle-point equation (i.e., the equation of motion) in a self-consistent manner:
\begin{align}
\frac{\delta E[U]}{\delta U} \bigg{|}_{U=U_{\mathrm{cl}}}=0.
\label{eq:me}
\end{align}
This approach is therefore referred to as the mean-field (or semiclassical) picture in the quantum field theory. The minimized energy from Eq.~\eqref{eq:me} is identified with the classical nucleon mass:
\begin{align}
M_{\mathrm{cl}} &= E[U_{\mathrm{cl}}].
\end{align}
The sum over the eigenenergies of the Dirac continuum~\eqref{eq:newf} contains a logarithmic ultraviolet (UV) divergence, which must be regularized. For further details, see Refs.~\cite{Diakonov:1996sr, Diakonov:1997vc}.

\subsection{Zero-mode quantization}
The classical nucleon is in fact degenerate in its spin/isospin quantum numbers and momentum. This degeneracy is lifted only after performing zero-mode quantization—specifically, flavor rotations and translations—of the mean field, which endows the baryon with definite quantum numbers. Moreover, the subleading spin-flavor structures in the $1/N_{c}$ expansion, essential for studying NLO chiral-odd GPDs, emerge as rotational and translational corrections to the mean field. Therefore, these zero modes must be treated explicitly. In this work, we consider translational modes to zeroth order and rotational modes to first order. While rotational corrections generate new spin-flavor structures, translational corrections provide only kinematical corrections. Consequently, we focus on the rotational zero mode.

The time-dependent pion field fluctuations around the mean-field solution involve rotational (and translational) zero modes. The rotational mode can be incorporated by introducing a unitary flavor rotation matrix $R(t)$ that rotates the mean field as
\begin{align}
U_{\mathrm{cl}} &\to R(t)U_{\mathrm{cl}}R^{\dagger}(t), 
\end{align}
which induces a shift in the Dirac operator,
\begin{align}
D(U_{\mathrm{cl}}) &\to R(t) \left[i\partial_{0}- H(U_{\mathrm{cl}}) -\Omega \right] R^{\dagger}(t),
\end{align}
where the angular velocity matrix $\Omega$ is defined as
\begin{align}
\Omega &\equiv \frac{1}{2}\Omega^{a} \tau^{a} = -\frac{i}{2} \mathrm{tr}[R^{\dagger}(t)\dot{R}(t) \tau^{a}] \tau^{a}, \cr
&\text{with} \quad \dot{R}(t)= \frac{d}{dt}R(t).
\label{eq:av}
\end{align}
The trace $\mathrm{tr}$ runs over flavor space. Having quantized these modes, corresponding to the canonical quantization rule $\Omega^{a} \to J^{a}/I$, one arrives at the Hamiltonian of a spherical top~\cite{Christov:1995vm},
\begin{align}
H_{\mathrm{rot}} = \frac{\bm{J}^{2}}{2I},
\label{eq:rot}
\end{align}
which serves as a correction to the classical nucleon mass. Here, $I$ denotes the moment of inertia (under rigid rotation), expressed as
\begin{align}
I = \frac{N_{c}}{6} \sum_{\substack{j,\mathrm{non} \\ n,\mathrm{occ}}} \frac{1}{E_{j}- E_{n}} \langle  n | \bm{\tau} | j  \rangle \cdot \langle  j | \bm{\tau} | n  \rangle.
\label{eq:momi}
\end{align}
This result is given by double sums over single-particle levels, describing particle-hole excitations of the fermionic states $\langle n | \to | j \rangle$ by $\Delta G=1$ and their recurrence to the states $| n \rangle$ (see Sec.~\ref{sec:sym} for details). This quantity~\eqref{eq:momi} is logarithmically divergent and must be regularized; “non” denotes the non-occupied single-particle states with $E_{j} > E_{\mathrm{lev}}$. 

Since the Wigner finite-rotation matrices are eigenfunctions of the Hamiltonian~\eqref{eq:rot} of spherical rigid rotors, the rotational wave functions for a baryon, which obey the ``hedgehog'' constraint, can be defined by these Wigner rotation matrices:
\begin{align}
\phi_{B}(R) &\equiv \phi^{S=T}_{S_{3}T_{3}}(R) \cr
&= (-1)^{T+T_{3}}\sqrt{2T+1}D^{T=S}_{-T_{3} S_{3}}(R),
\label{eq:sf_wf}
\end{align}
where the collective representation $B$ of the baryon quantum numbers indicates
\begin{align}
B= \{ S=T,S_{3},T_{3} \}.
\label{eq:bary_qn}
\end{align}
Here, $S_{3}$ and $T_{3}$ represent the projections of the spin and isospin quantum numbers. The normalization of Eq.~\eqref{eq:sf_wf} is chosen so that the matrix element of the rotational wave functions is normalized to the identity, with the integration measure normalized as $\int dR = 1$:
\begin{subequations}
\begin{align}
\int dR \, \phi^{*}_{B'}(R) \phi_{B}(R) = \delta_{B'B}, \\
\delta_{B'B} \equiv \delta_{S'S}\delta_{T'T}\delta_{S_{3}'S_{3}}\delta_{T_{3}'T_{3}}. 
\end{align}
\end{subequations}
The lowest-lying baryons corresponding to the representations with $S=T=1/2$ and $S=T=3/2$ appear as the nucleon and the $\Delta$ baryon, respectively. The mass splitting between them is computed by taking the matrix element of the rotational Hamiltonian~\eqref{eq:rot} between the rotational wave functions~\eqref{eq:sf_wf}:
\begin{align}
M_{\Delta} - M_{N} = \frac{3}{2I} = \mathcal{O}(N^{-1}_{c}).
\label{eq:correction}
\end{align}
The $N_c$ scaling of this mass splitting is two powers lower than that of the classical nucleon mass~\eqref{eq:te}, and thus appears as a mere kinematical correction to the observables. Therefore, in the following, we consider only the rotational corrections that generate nontrivial spin-flavor structures (i.e., those carrying new dynamical information), while neglecting the kinematical corrections~\eqref{eq:correction}.

\section{Chiral-odd GPDs in mean-field picture}
\subsection{Matrix element of QCD operator}
The study of nucleon matrix elements of QCD operators can be carried out using a method based on the mean-field picture of baryons in the large-$N_{c}$ limit. To this end, we first need to find an effective operator corresponding to the QCD operator of interest. The non-local leading-twist QCD operators $(\Gamma = \gamma^{+}, \gamma^{+}\gamma_{5}, i\sigma^{+j})$ we consider are defined as
\begin{align}
O^{f}(z) =  \bar{\psi}^{f}(-z/2) \left[-z/2, z/2\right] \Gamma \psi^{f}(z/2).
\label{eq:op}
\end{align}
where $f$ denotes the flavor components, $f = (u, d)$. This QCD operator should be matched to the scale of the effective theory at $\mu \approx \bar{\rho}^{-1}$ and rewritten in terms of the effective degrees of freedom. For higher-twist operators, gluon contributions $\sim (M\bar{\rho})^{0}$ are not parametrically  suppressed and must be represented through suitable effective operators (see Refs.~\cite{Diakonov:1995qy, Balla:1997hf, Polyakov:1998ip, Polyakov:1996kh, Dressler:1999zi, Weiss:2021kpt, Polyakov:2018exb, Kim:2023pll}). On the other hand, Ref.~\cite{Balla:1997hf} demonstrated that the gluon contribution to the leading-twist non-local operator~\eqref{eq:op} is suppressed at the low normalization point due to the diluteness $\sim (M\bar{\rho})^{2}$ of instantons, so that only the pure derivative terms—originating from the non-locality between quark fields—survive. As a result, the contribution of the gauge connection $[\ldots]$ can be safely neglected within the parametric accuracy of the present work, and the corresponding effective operator in Euclidean space ($\bar{\psi} \to -i\psi^{\dagger}$) is given by
\begin{align}
O^{f}_{\mathrm{eff}}(z) = - i\psi^{\dagger}_{\alpha h}(-z/2)  ( \Gamma \tau^{f} )^{h}_{g} \psi^{\alpha g}(z/2).
\label{eq:eff}
\end{align}
where the SU(2) flavor matrix $\tau^{f}$ denotes either isoscalar or isovector components:
\begin{align}
\tau^{f} = \left\{\begin{array}{c r} \bm{1}  &(f=u+d) \\[2ex] \tau^{3}  &(f=u-d) \end{array} \right\}.
\end{align}

Next, the baryon states in the momentum representation are defined by the Fourier transform of the Ioffe-type baryon current:
\begin{align}
| B, \bm{p} \rangle &= \lim_{y^{0} \to \infty} \sqrt{2M_{N}} \int d^{3} y  e^{i y\cdot p } J^{\dagger}_{B}(y) | 0 \rangle, \cr
\langle B', \bm{p}'|  &= \lim_{y'^{0} \to -\infty} \sqrt{2M_{N}} \int d^{3} y'  e^{-i y'\cdot p' } \langle 0 | J_{B^{\prime}}(y'),
\label{eq:N_states}
\end{align}
where the initial and final nucleon states are at the infinite Euclidean time separation, and the normalization of the baryon states in the the large-$N_{c}$ limit is $\sqrt{2M_{N}}$. $J_{B}$ is the Ioffe-type baryon current, defined as
\begin{subequations}
\label{eq:Ioffe}
\begin{align}
J_{B^{\prime}}(y^{\prime})&=\frac{1}{N_{c}!} \Lambda^{h_1\ldots h_{N_c}}_{B^{\prime}} \epsilon^{\alpha_{1}\ldots\alpha_{N_c}} \cr
&\times \psi_{\alpha_1 h_1} (y^{\prime}) \ldots \psi_{\alpha_{N_{c}} h_{N_c}} (y^{\prime}), \\[1ex]
J^{\dagger}_{B}(y)&=\frac{1}{N_{c}!} (\Lambda^{g_1\ldots g_{N_c}}_{B})^{*} \epsilon^{\beta_{1}\ldots\beta_{N_c}} \cr
&\hspace{-0.5cm}\times (- i  \psi^{\dagger}(y) \gamma_{4})_{\beta_1 g_1} \ldots (- i\psi^{\dagger}(y) \gamma_{4})_{\beta_{N_{c}} g_{N_c}}.
\end{align}
\end{subequations}
The baryon current~\eqref{eq:Ioffe} consists of $N_{c}$ quark fields in a totally antisymmetric color-singlet configuration. The color indices are denoted by $\alpha, \beta$. The spin-flavor indices $h, g$ of the quarks are contracted with the baryon spin-flavor matrix $\Lambda_{B}$, which carries the quantum numbers of a given baryon. Furthermore, in the $1/N_{c}$ expansion of the matrix element between baryon states, the three-momentum and energies of the baryon states $B'$ and $B$ have the following $N_{c}$ scalings
\begin{align}
&|\bm{p}'|,|\bm{p}| = \mathcal{O}(N^{0}), \nonumber \\[1ex]
&p^{\prime 0},p^{0} = M_{N} + \mathcal{O}(N^{-1}_{c}) = \mathcal{O}(N_c).
\label{eq:3_mo}
\end{align}
Under the Wick rotation $-ip_{4} = p_{0}$, the Euclidean momentum is understood as the nucleon mass
\begin{align}
-i p'_{4} = -i p_{4} = M_{N} + \mathcal{O}(N^{-1}_{c}).
\end{align}

\begin{figure}[t]
\centering
\includegraphics[scale=0.23]{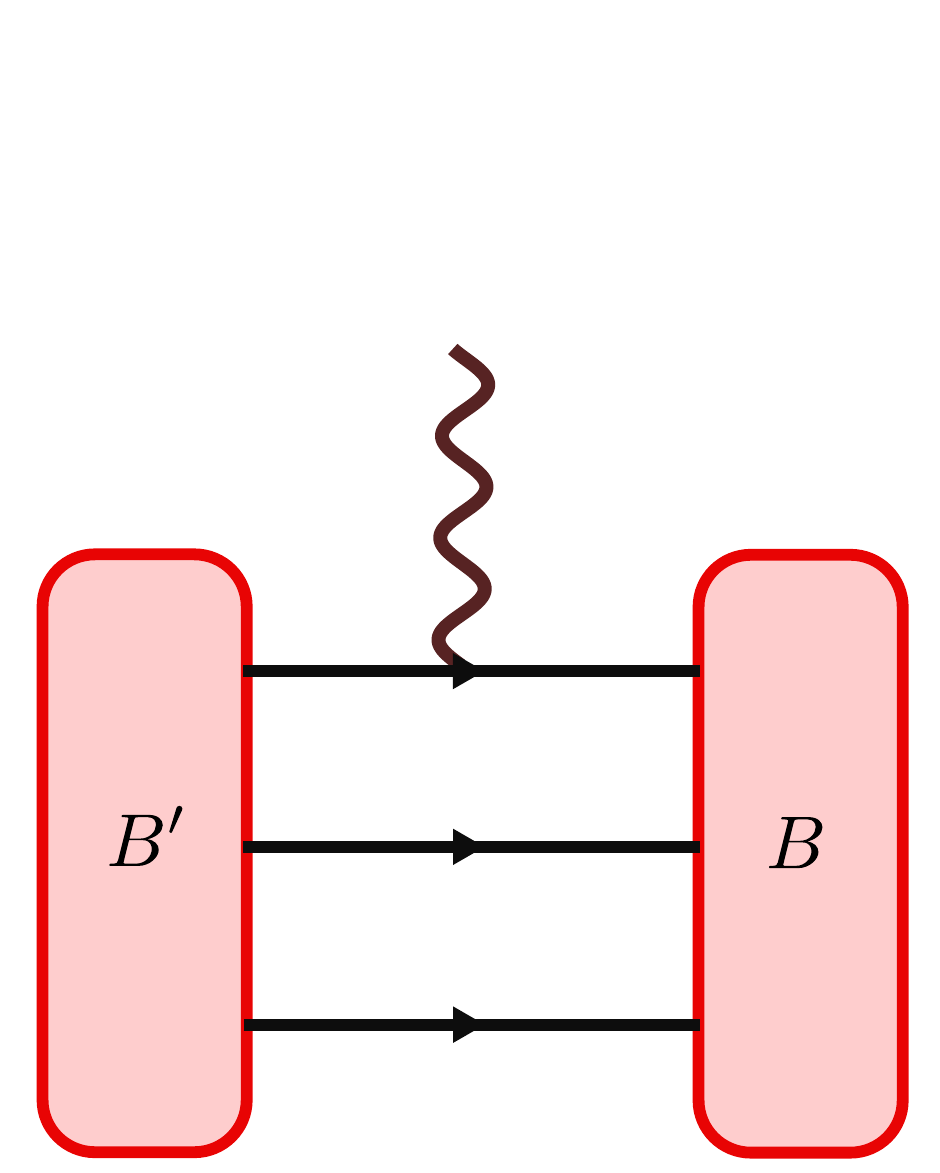} \hspace{0.3cm}
\includegraphics[scale=0.23]{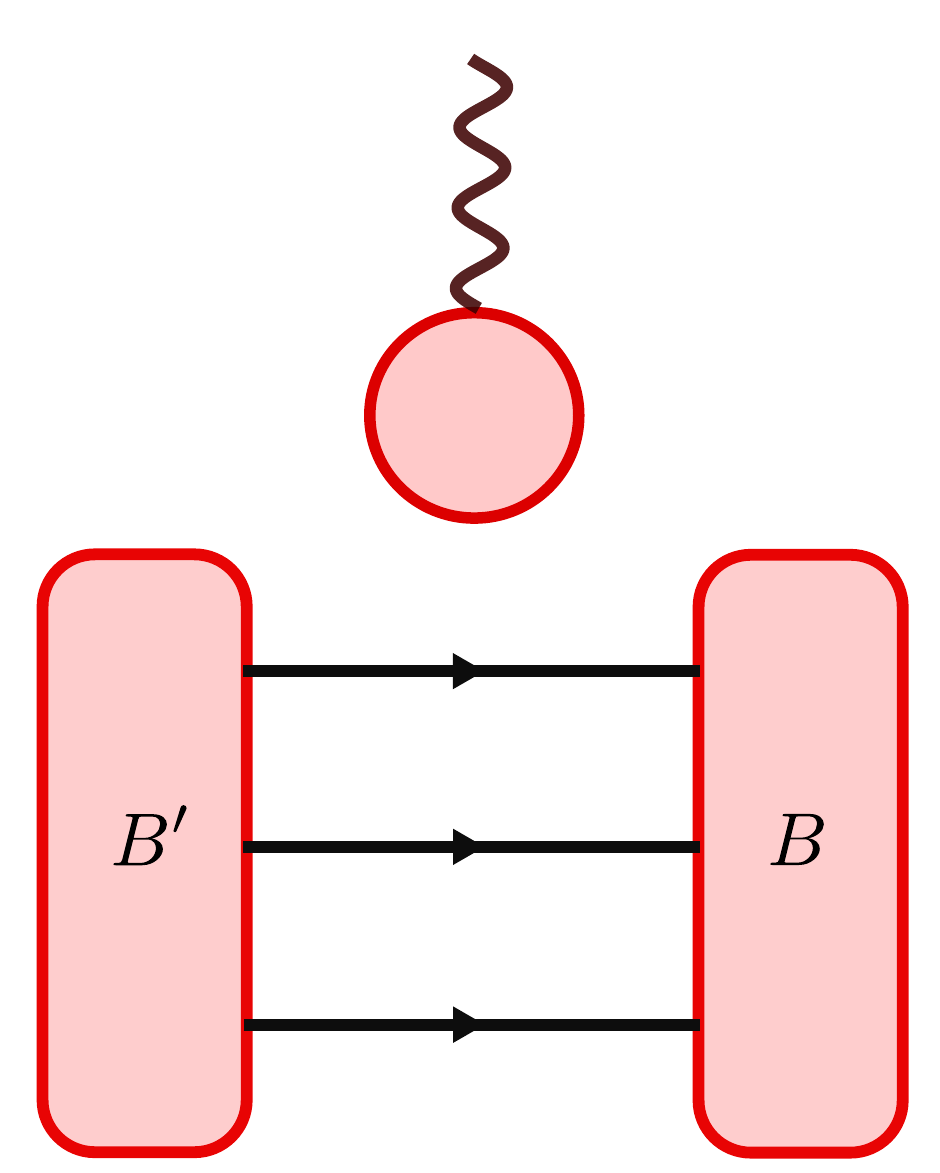}
\caption{Connected (left panel) and disconnected (right panel) diagrams in the backgound of the pion mean-field}
\label{fig:dia}
\end{figure}

Having performed all the contractions of the quark fields between the initial and final baryon states~\eqref{eq:Ioffe}, considered both the rotational and translational zero modes of the pion mean field, and taken into account the $N_{c}$ scalings of the 4-momentum~\eqref{eq:3_mo}, we find that the overlap of the baryon states~\eqref{eq:N_states} yields the correct normalization of the baryon states in the large-$N_{c}$ limit:
\begin{align}
\langle B', \bm{p}' | B, \bm{p} \rangle = 2M_{N}(2\pi)^{3}\delta^{(3)}(\bm{p}'-\bm{p}) \delta_{B'B}.
\end{align}
Similarly, by sandwiching the effective operator~\eqref{eq:eff} between the baryon states~\eqref{eq:N_states}, we compute the matrix element using functional integrals:
\begin{align}
&\langle B',\bm{p}' |  O^{f}(z) | B,\bm{p} \rangle = 2M_{N}   \lim_{\substack{y'_{4} \to \infty \\ y_{4} \to -\infty}} \frac{1}{Z_{\mathrm{eff}}} \cr
&\times e^{i y_{4} p_{4} } e^{-i y'_{4} p'_{4} }   \int d^{3} y  d^{3} y' \, e^{i \bm{y}\cdot \bm{p} } e^{-i \bm{y}'\cdot \bm{p}' }   \cr
&\times\int \mathcal{D}U   \mathcal{D}\psi^{\dagger}  \mathcal{D}\psi J_{B'}(\bm{y}',y'_{4}) O^{f}_{\mathrm{eff}}(z) J^{\dagger}_{B}(\bm{y},y_{4}) \cr
&\times \exp\left[ \int d^{4}x \, \psi^{\dagger}_{\alpha h}(x) (i \slashed{\partial} + i M U^{\gamma_{5}})^{h}_{g} \psi^{\alpha g}(x)\right].
\label{eq:3corr}
\end{align}
We first perform the contractions between the quark fields in the baryon currents, including those in the effective operator $O_{\mathrm{eff}}$. There are two distinct contributions (see Fig.~\ref{fig:dia}). One arises from contractions between the initial/final baryon currents and the effective operator $O_{\mathrm{eff}}$ (connected diagram). The other involves contractions performed solely between the initial and final baryon currents, with the bilinear operator contracting with itself (disconnected diagram). In both cases, $N_{c} + 1$ quark propagators appear, implying that the pre-exponential part in Eq.~\eqref{eq:3corr} becomes independent of the quark fields. Thus, the functional integrals over the quark fields [cf.~\eqref{eq:action}] can be performed exactly, leaving only the functional integral over $U$. All $N_{c} + 1$ quark propagators are then expressed in terms of the single-particle wave functions~\eqref{eq:newf} using the spectral representation (refer to \cite{Christov:1995vm}).

\subsection{Spin-flavor structure \label{sec:sfs}}
Having performed the functional integral over $U$ and taken into account the zero modes of the pion mean-field in Eq.~\eqref{eq:3corr}, we finally obtain the results for the baryon matrix element of the effective operator and for the corresponding spin-flavor structure for both the LO and NLO contributions in the $1/N_{c}$ expansion.

{\it{Leading order.}} Considering the zeroth order in rotational and translational corrections, we derive the leading-order contributions in the $1/N_{c}$ expansion (see Ref.~\cite{Kim:2024ibz} for details):
\begin{align}
&
\left.
\begin{array}{l}
\mathcal{M}^{u+d}_{\mathrm{GPDs}}[\Gamma]
\\[2ex]
\mathcal{M}^{u-d}_{\mathrm{GPDs}}[\Gamma]
\end{array}
\right\} 
\overset{\text{LO}}{=} 2 M^{2}_{N} N_{c}
\left\{
\begin{array}{c}
\langle \bm{1} \rangle
\\[2ex]
\langle D^{3k} \rangle
\end{array}
\right\}
\nonumber \\[2ex]
& \times \sum_{n,\mathrm{occ}} \;
\int \frac{dz^{0}}{2\pi}  e^{i ( xM_{N}-E_{n} ) z^{0}  } 
\nonumber \\
&\times \langle n |
\; \gamma^{0} \Gamma
e^{-i \frac{\hat{p}^{3}}{2}z^{0}} e^{i\bm{\Delta} \cdot \bm{\hat{X}}}\left\{
\begin{array}{c}
1
\\[1ex]
\tau^{k}
\end{array}
\right\}
e^{-i \frac{\hat{p}^{3}}{2}z^{0}}|n\rangle,
\label{eq:general}
\end{align}
To facilitate comparison with the standard representation of the matrix element, we now express the present results in Minkowski space. The constraints $z^{+} = 0$ and $\bm{z}_{\perp} = 0$ in Eq.~\eqref{eq:General_ME} imply
\begin{align}
z^{0} = - z^{3}, \quad \bm{z}_{\perp}=0.
\end{align}
The operator function $e^{i \bm{\Delta} \cdot \hat{\bm{X}}}$  in Eq.~\eqref{eq:general} originates from the quantization of the translational motion of the mean field. Here, $\bm{\hat{p}}$ and $\bm{\hat{X}}$ denote the single-particle momentum and displacement operators, respectively, and satisfy the canonical commutation relation
\begin{align}
[\hat{p}^{i}, \hat{X}^{j}] = - i \delta^{ij}.
\end{align}
The notation $\langle \dots \rangle$ denotes the integration over collective flavor rotations, evaluated with the spin-flavor wave functions~\eqref{eq:sf_wf}, and is defined as
\begin{align}
&\langle ... \rangle \equiv \int dR \, \phi_{S'_3T'_3}^{*S'=T'}(R) \, ... \, \phi_{S_3T_3}^{S=T}(R).
\label{eq:collect_integral}
\end{align}

The explicit expressions for the diagonal matrix elements ($S'=S$ and $T'=T$) of the spin-flavor operator are written in terms of multipole spin operators:
\begin{subequations}
\label{eq:sf_me_1}
\begin{align}
\langle \bm{1} \rangle &= \delta_{B^{\prime} B},  \\
 \langle D^{3i} \rangle 
&= - \frac{J^{i}_{S'_{3} S_{3}}}{\sqrt{S(S+1)}} \langle T T_{3}, 1 0 | T T'_{3} \rangle.
\end{align}
\end{subequations}
where the matrix element of the spin operator can be expressed in terms of SU(2) Clebsch-Gordan coefficients in the spherical basis as
\begin{align}
J^{a}_{S'_{3} S_{3}} &= \sqrt{S(S+1)}\langle S S_{3}, 1 a | S S'_{3} \rangle,  \\[1ex]
 &\mathrm{with} \quad (a=0,\pm1; \  S'_{3},S_{3}=0, \cdot\cdot\cdot,\pm S). \nonumber
\end{align}
As shown in Eqs.~\eqref{eq:general} and \eqref{eq:sf_me_1}, each flavor component of the LO matrix element contains only a specific spin multipole structure. For instance, the flavor-singlet (non-singlet) component involves exclusively the monopole (dipole) spin structure. This indicates that NLO contributions are required to account for the dipole (monopole) spin structure in the flavor-singlet (non-singlet) matrix element. 

{\it{Next-to-leading order.}}  
The NLO contribution in the $1/N_{c}$ expansion is obtained by expanding the matrix element~\eqref{eq:3corr} with respect to the angular velocity~\eqref{eq:av}, including the effects of the time non-locality of the effective operator, and by taking the subleading terms (see Ref.~\cite{Pobylitsa:1998tk, Schweitzer2003,Ossmann:2004bp} for details):
\begin{align}
&
\left.
\begin{array}{l}
\mathcal{M}^{u+d}_{\mathrm{GPDs}}[\Gamma]
\\[2ex]
\mathcal{M}^{u-d}_{\mathrm{GPDs}}[\Gamma]
\end{array}
\right\}
\overset{\text{NLO}}{=} \frac{M^{2}_{N} N_{c}}{I}
\left\{
\begin{array}{c}
\langle J^{k} \rangle
\\[2ex]
\frac{1}{2} \langle \{J^{k},D^{3l}\} \rangle
\end{array}
\right\} 
\nonumber \\[2ex]
& \times \int \frac{dz^{0}}{2\pi} \bigg{[} \sum_{\substack{n,\mathrm{occ} \\ j,\mathrm{all}}} \frac{e^{-i E_{n} z^{0}}}{E_{n}-E_{j}}  - \sum_{\substack{n,\mathrm{all} \\ j,\mathrm{occ}}} \frac{e^{-i E_{j}z^{0}}}{E_{n}-E_{j}}    \cr
&+  \frac{d}{dx M_{N}} \sum_{\substack{n,\mathrm{occ} \\ j,\mathrm{all}}} e^{- i E_{n}  z^{0} } \bigg{]}
  e^{i  xM_{N} z^{0} } 
\nonumber \\
&\times \langle n | \tau^{k} | j \rangle \langle j |
\; \gamma^{0} \Gamma
e^{-i \frac{\hat{p}^{3}}{2} z^{0}} e^{i\bm{\Delta} \cdot \bm{\hat{X}}}\left\{
\begin{array}{c}
1
\\[1ex]
\tau^{l}
\end{array}
\right\}
e^{-i \frac{\hat{p}^{3}}{2} z^{0}}|n\rangle, \cr
&+..., 
\label{eq:general2}
\end{align}
where the ellipsis indicates new dynamical contributions containing the same spin-flavor structures as those in the LO matrix element~\eqref{eq:general}, which provide corrections to Eq.~\eqref{eq:general} but will not be considered in this work. Due to the rotational correction, new spin-flavor operators emerge in Eq.~\eqref{eq:general2}. In particular, the collective spin operator $\bm{J}$ in the NLO matrix element accompanies the dynamical parameter $I = O(N_c)$ [see Eq.~\eqref{eq:rot}], leading to a reduction in the $N_c$ scaling. Therefore, the $N_c$ scaling of the NLO matrix element is suppressed by one power compared to that of the LO matrix element.

The diagonal matrix elements of the spin-flavor operator are expressed in terms of multipole spin operators:
\begin{subequations}
\label{eq:quad}
\begin{align}
\langle J^{k} \rangle &= J^{k}_{S^{\prime}_{3} S_{3}} \delta_{T_{3}'T_{3}}, \\[1ex]
\langle \{J^{k}, D^{3l} \} \rangle
&= - \left[\frac{Q^{kl}_{S^{\prime}_{3} S_{3}} + \frac{2}{3} J^{2}_{S^{\prime}_{3} S_{3}} \delta^{kl}}{\sqrt{S(S+1)}} \right] \nonumber \\[1ex]
&\times\langle T T_{3}, 1 0 | T T'_{3} \rangle.
\end{align}
\end{subequations}
By inserting Eq.~\eqref{eq:quad} into Eq.~\eqref{eq:general2}, one clearly sees that at NLO the flavor-singlet and flavor-non-singlet matrix elements contain, respectively, the dipole spin operator and the monopole and quadrupole spin operators. These contributions complement the missing spin structures in the LO matrix element. 

Furthermore, the symmetrized and traceless higher-multipole (quadrupole) spin operator, defined as
\begin{align}
Q^{kl}_{S'_{3}S_{3}} &= \frac{1}{2}\left( J^{k}J^{l} + J^{l}J^{k} - \frac{2}{3}S(S+1)\delta^{kl}\right)_{S'_{3}S_{3}},
\end{align}
emerges from the NLO spin-flavor operator~\eqref{eq:quad}. Since the nucleon matrix element can produce only up to a dipole spin structure, the quadrupole projection of this operator vanishes and is therefore irrelevant to this work. However, this spin structure becomes essential in the study of higher-spin particles, such as the $\Delta$ baryon~\cite{Fu:2022bpf, Fu:2024kfx} and the $N \to \Delta$ transition matrix element~\cite{Kroll:2022roq, Diehl:2024bmd, Kroll:2025osx}.

\subsection{Large-$N_{c}$ kinematics \label{sec:lnck}}
Before discussing how the chiral-odd GPDs are derived from Eq.~\eqref{eq:general2}, we first discuss the kinematic variables within the framework of the $1/N_c$ expansion. In the large-$N_c$ limit, the $N_c$ scaling of the kinematic variables can be straightforwardly determined from Eq.~\eqref{eq:3_mo}. Specifically, the $N_c$ scalings of the average and difference of the initial and final momenta are
\begin{align}
P^{i}&= \mathcal{O}(N^{0}_{c}),  &&P^{0}= M_{N} + \mathcal{O}(N^{-1}_{c}), \cr
\Delta^{i}&=\mathcal{O}(N^{0}_{c}),   &&\Delta^{0}= \mathcal{O}(N^{-1}_{c}).
\label{eq:nc_mo}
\end{align}
By choosing the 3-axis as the direction of the light-cone spatial axis, the components of the four-momentum can be related to the light-cone variables in the large-$N_c$ limit as follows:
\begin{align}
&P^{+} = P^{-} = \frac{M_{N}}{\sqrt{2}}, &&\Delta^{+} = -\Delta^{-} = \frac{\Delta^{3}}{\sqrt{2}}.
\label{eq:LC_var}
\end{align}
Using Eq.~\eqref{eq:LC_var}, the longitudinal momentum transfer $\xi$ (skewness) between the baryon states is defined by the $z$-component of the momentum transfer $\Delta^{3}$:
\begin{align}
\xi = -\frac{\Delta^{3}}{2M_{N}} = \mathcal{O}(N^{-1}_{c}).
\label{eq:skew}
\end{align}
Another kinematic variable relevant to the GPDs is the squared momentum transfer $t$, which can be expressed in terms of the squared transverse momentum transfer $|\bm{\Delta}_{\perp}|^{2}$ and the skewness variable~\eqref{eq:skew}:
\begin{align}
t & = -|\bm{\Delta}_{\perp}|^{2} - (2 \xi M_{N})^{2} = \mathcal{O}(N^{0}_{c}).
\label{eq:hierarchy2}
\end{align}
Lastly, the partonic variable $x$ is taken to be of order
\begin{align}
x = \mathcal{O}(N^{-1}_{c}),
\label{eq:p_3}
\end{align}
which corresponds to the standard regime considered in the $1/N_c$ expansion of nucleon parton distributions~\cite{Diakonov:1996sr}. 

Consequently, using the relations~\eqref{eq:skew} and~\eqref{eq:hierarchy2}, the matrix elements in Eqs.~\eqref{eq:general} and \eqref{eq:general2} can be re-expressed in terms of three independent GPD variables:
\begin{align}
\mathcal{M}_{\mathrm{GPDs}}[\Gamma] &= \text{function} (x,\bm{\Delta}; B',B) \nonumber  \\[1ex]
&\to \text{function} (x,\xi,t, \theta_{\Delta} ; B',B),
\label{eq:mul_g}
\end{align}
where $\theta_{\Delta}$ denotes the angular orientation of the transverse momentum transfer $\bm{\Delta}_{\perp}$. This angular dependence gives rise to the two-dimensional~(2D) multipole structure of the matrix element. In the following section, we will discuss this 2D multipole expansion.

\subsection{Multipole expansion}
Using the chiral-odd operator $\Gamma = i\sigma^{+j}$ and taking the proton quantum numbers $S = T = 1/2$ for the baryon states $B$ and $B'$ in Eq.~\eqref{eq:general2}, we derive the proton matrix element of the chiral-odd operator. We then expand this matrix element~\eqref{eq:mul_g} in terms of multipoles with respect to the 2D momentum transfer angle $\theta_{\Delta}$ and the multipole spin~\eqref{eq:quad}. Since the LO results in the $1/N_c$ expansion have been thoroughly studied in Ref.~\cite{Kim:2024ibz}, we focus here on the NLO contributions. The NLO results for the multipole expansion are obtained as follows:
\begin{subequations}
\label{eq:model_spin_flavor_2}
\begin{align}
\mathcal{M}^{u-d}_{\mathrm{GPDs}}[i\sigma^{+j}] = & \frac{2 M_N}{\sqrt{2}} \left[ \bm{1} \, X^{j}_{1} \, 
\frac{|\bm{\Delta}_{\perp}|}{2  M_N} Z_{\textrm{mf}, 1} \right], \label{eq:model_spin_flavor_2a}
\\ 
\mathcal{M}^{u+d}_{\mathrm{GPDs}}[i\sigma^{+j}] = & \frac{2 M_N}{\sqrt{2}}
\biggl{[} i \epsilon^{3jm} \sigma^{m} \, X_0  \,
Z_{\textrm{mf}, 0} 
\nonumber \\
&+i \epsilon^{3jm} \sigma^{3} \, X^{m}_{1} \,
\frac{|\bm{\Delta}_{\perp}|}{2  M_N} \,
\tilde{Z}_{\textrm{mf}, 1}
\nonumber \\
&+ i \epsilon^{3jl} \sigma^{m} \, X^{lm}_2 \,
\frac{|\bm{\Delta}_{\perp}|^{2}}{4M_N^2} \,
Z_{\textrm{mf}, 2} \biggr{]}, \label{eq:model_spin_flavor_2b}
\end{align}
\end{subequations}
where the irreducible rank-$n$ tensors in 2D space are defined as
\begin{align}
X_{0}&\equiv1, \quad X^{i}_{1}\equiv\frac{\Delta^{i}_{\perp}}{|\bm{\Delta}_{\perp}|}, \quad X^{ij}_{2}\equiv\frac{\Delta^{i}_{\perp}\Delta^{j}_{\perp}}{|\bm{\Delta}_{\perp}|^{2}} - \frac{1}{2}\delta^{ij},
\end{align}
with $i,j = 1,2$. We also use the following shorthand notation for the spin structure:
\begin{align}
\bm{1} \equiv \delta_{S'_{3}S_{3}}, \quad \frac{\bm{\sigma}}{2} \equiv   \bm{J}_{S'_{3}S_{3}}.
\end{align}
The quantities $Z_{\mathrm{mf},n}$ represent the multipole (mean-field) GPDs and emergies at NLO in the $1/N_{c}$ expansion. The subscript $n$ denotes the multipole rank corresponding to the 2D momentum transfer. Similar to the standard chiral-odd GPDs, the mean-field GPDs are functions of the partonic variables:
\begin{align}
Z_{\mathrm{mf}} = \text{function}(N_{c}x, N_{c}\xi, t).
\label{eq:model_spin_flavor_3}
\end{align}
Their explicit expressions are presented in Sec.~\ref{sec:mfgpds}.

In the present large-$N_c$ analysis, we find that the NLO mean-field GPDs exhibit the following $N_c$ scalings [cf.~\eqref{eq:mfgpds_1}]:
\begin{align}
& \{Z_{\mathrm{mf},0}, \, Z_{\mathrm{mf},1}, \, \tilde Z_{\mathrm{mf},1}, \, Z_{\mathrm{mf},2} \}(x,\xi,t)
\nonumber \\[.5ex]
& \sim  \{N^{1}_{c}, \, N^{2}_{c}, \, N^{2}_{c}, \, N^{3}_{c} \}
\times \mathrm{function}(N_{c}x,N_{c}\xi,t),
\label{eq:Nc_zero_functions}
\end{align}
which is consistent with the results of Ref.~\cite{Kim:2024ibz}. By combining the known $N_c$ scalings of the LO mean-field GPDs~\cite{Kim:2024ibz} with the present NLO analysis~\eqref{eq:Nc_zero_functions}, we conclude that the large-$N_c$ behavior of the individual chiral-odd GPDs is given by
\begin{subequations}
\label{eq:Ncscaling}
\begin{align}
& \{H^{u-d}_{T}, \, \tilde{H}^{u-d}_{T}, \, E^{u-d}_{T}, \, \tilde{E}^{u-d}_{T}\}(x,\xi,t)
\nonumber \\[.5ex]
& \sim \ \{ N^{2}_{c}, \, N^{4}_{c}, \, N^{4}_{c}, \, N^{3}_{c} \} \times \mathrm{function}(N_{c}x,N_{c}\xi,t),  \\[1ex]
& \{H^{u+d}_{T}, \, \tilde{H}^{u+d}_{T}, \, E^{u+d}_{T}, \, \tilde{E}^{u+d}_{T} \}(x,\xi,t)
\nonumber \\[.5ex]
& \sim \ \{N_{c}, \, N^{3}_{c}, \, N^{3}_{c}, \, N^{2}_{c} \} \times \mathrm{function}(N_{c}x,N_{c}\xi,t),
\end{align}
\end{subequations}
along with the nontrivial large-$N_c$ relation
\begin{align}
2\tilde{H}^{u-d}_{T}(x, \xi, t) = - E^{u-d}_{T}(x, \xi, t).
\label{eq:largeNc_relation_1}
\end{align}

Taking Eqs.~\eqref{eq:Nc_zero_functions}, \eqref{eq:Ncscaling}, and \eqref{eq:largeNc_relation_1}, along with the large-$N_c$ kinematics discussed in Sec.~\ref{sec:lnck}, into consideration, the mean-field GPDs~\eqref{eq:model_spin_flavor_2} can be related to the standard chiral-odd GPDs~\eqref{eq:General_ME_GPDs} (see Ref.~\cite{Kim:2024ibz} for details) as follows:
\begin{subequations}
\label{eq:largeNc_relation_nonsinglet}
\begin{align}
H^{u+d}_{T} + \left( \frac{t}{8M^{2}_{N}} - \frac{\xi^2}{2}  \right) E^{u+d}_{T} + \xi \tilde{E}^{u+d}_{T}
&= Z_{\mathrm{mf},0},
\label{relation_0}
\\
H^{u-d}_{T} + 2 \tilde{H}^{u-d}_{T} + E^{u-d}_{T}
&= Z_{\mathrm{mf},1},
\label{relation_1}
\\[1ex]
 - \xi E^{u+d}_{T} + \tilde{E}^{u+d}_{T} 
&= \tilde{Z}_{\mathrm{mf},1},
\label{relation_1tilde}
\\[1ex]
  E^{u+d}_{T}
&= Z_{\mathrm{mf},2}.
\label{relation_2}
\end{align}
\end{subequations}
These relations are homogeneous in $N_c$ and offer a more transparent physical interpretation than the covariant expressions of the chiral-odd GPDs. A similar set of combinations, optimized for the SO(3) partial wave expansion in the cross channel, was constructed in Ref.~\cite{Pire:2014fwa}. It would be interesting to investigate the connection between those combinations and the present multipole structures given in Eq.~\eqref{eq:largeNc_relation_nonsinglet}. In what follows, we adopt the mean-field (multipole) GPDs as the basis of our analysis, including the proof of polynomiality and the derivation of associated sum rules. 

\subsection{Mean-field GPDs \label{sec:mfgpds}}
The proton mean-field GPDs defined in Eqs.~\eqref{eq:model_spin_flavor_2} and \eqref{eq:model_spin_flavor_3} are derived in the single-particle representation:
\begin{align}
&Z_{\mathrm{mf}}=   \frac{ M_{N} N_{c}}{8 I}  \int \frac{dz^{0}}{ 2\pi} \cr
&\times \bigg{[}\sum_{ \substack{n,\mathrm{occ} \\ j,\mathrm{all}} } \frac{e^{-i E_{n} z^{0} }}{E_{n} - E_{j}} -\sum_{ \substack{n,\mathrm{all} \\ j,\mathrm{occ}} } \frac{e^{-i E_{j} z^{0} }}{E_{n} - E_{j}} \cr
 &+ \frac{d}{dx M_{N}} \sum_{ \substack{n,\mathrm{occ} \\ j,\mathrm{all}} } e^{-i  E_{n} z^{0}}  \bigg{]} e^{i  x M_{N} z^{0} } \cr
&\times  \langle n | \tau^{b} | j \rangle \langle j | O^{b}  e^{-i \frac{\hat{p}^{3}}{2} z^{0}} e^{i \bm{\Delta} \cdot \bm{\hat{X}}} e^{-i  \frac{\hat{p}^{3}}{2} z^{0}} | n \rangle, 
\label{eq:mfgpds_1}
\end{align}
where the single-particle spin-isospin operators $O^{b}$ corresponding to each mean-field GPD are defined as
\begin{align}
O^{b} &= i \epsilon^{3ab}  (1+\gamma^{0}\gamma^{3})\gamma^{a} &&\left(\text{for} \ Z_{\mathrm{mf},0}\right), \cr
&= \frac{4}{3} \frac{\Delta^{a}_{\perp}}{|\bm{\Delta}_{\perp}|^{2}}  (1+\gamma^{0}\gamma^{3})\gamma^{a} \tau^{b} && \left( \ \frac{Z_{\mathrm{mf},1}}{M_{N}}\right), \cr
&= 4 i \epsilon^{3ac}  \frac{\Delta^{c}_{\perp}}{|\bm{\Delta}_{\perp}|^{2}} \delta^{b3} (1+\gamma^{0}\gamma^{3})\gamma^{a} &&\left( \ \frac{\tilde{Z}_{\mathrm{mf},1}}{M_{N}}\right), \cr
&= 16  i \epsilon^{3ac} \frac{1}{|\bm{\Delta}_{\perp}|^{4}} (\Delta^{b}_{\perp}\Delta^{c}_{\perp} - \frac{1}{2} |\bm{\Delta}_{\perp}|^{2} \delta^{bc})  \cr
&\hspace{0.5cm}\times (1+\gamma^{0}\gamma^{3})\gamma^{a} &&\left( \ \frac{Z_{\mathrm{mf},2}}{M^{2}_{N}}\right). 
\label{eq:mfgpds_2}
\end{align}
The single-particle operators $O^{b}$ reflect the characteristics of the multipole structure of each mean-field GPD. As discussed in Eqs.~\eqref{eq:skew} and \eqref{eq:hierarchy2}, the three-momentum transfer $\bm{\Delta}$ appearing in Eqs.~\eqref{eq:mfgpds_1} and \eqref{eq:mfgpds_2} can be related to the GPD variables.

\section{Moments in mean-field picture \label{sec:mimfp}}
The moments of the chiral-odd GPDs exhibit polynomiality in $\xi$ and are related to generalized tensor form factors. We investigate these features using the mean-field/multipole GPD basis defined in Eq.~\eqref{eq:largeNc_relation_nonsinglet}. The $m$-th moments of the chiral-odd GPDs can be obtained from Eq.~\eqref{eq:mfgpds_1} as
\begin{align}
M^{(m)}(\xi,t)=\int dx \, x^{m-1} Z_{\mathrm{mf}},
\label{eq:moment_0}
\end{align}
which leads to
\begin{align}
&M^{(m)}(\xi,t) \cr
&= \frac{N_{c}}{8IM^{m-1}_{N}} \Bigg{[}\sum_{k=0}^{m-1} \binom{m-1}{k} \sum^{k}_{i=0} \binom{k}{i} \frac{1}{2^{k}} \cr
&\hspace{0.7cm}\times \bigg{\{} \sum_{\substack{j,\mathrm{all} \\ n,\mathrm{occ} }} E^{m-k-1}_{n} - \sum_{\substack{j,\mathrm{occ} \\ n,\mathrm{all} }} E^{m-k-1}_{j} \bigg{\}} \frac{A^{ki}}{E_{n} - E_{j}} \cr
&-(m-1) \sum_{k=0}^{m-2} \binom{m-2}{k} \sum^{k}_{i=0} \binom{k}{i}   \sum_{{n,\mathrm{occ} }} \frac{E^{m-k-2}_{n}}{2^{k}}  B^{ki} \Bigg{]},
\label{eq:moment}
\end{align}
where $A\equiv A(\xi,t)$ and $B\equiv B(\xi,t)$ represent matrix elements involving single and double sums over single-particle levels, respectively. They are defined as:
\begin{subequations}
\label{eq:moment_1}
\begin{align}
A^{ki}(\xi,t)&\equiv \langle n | \tau^{b} | j \rangle \langle j | O^{b} (\hat{p}^{3})^{k-i}   e^{i \bm{\Delta} \cdot \bm{\hat{X}}} (\hat{p}^{3})^{i} | n \rangle,  \label{eq:moment_1a} \\[.5ex]
B^{ki}(\xi,t)&\equiv \langle n | \tau^{b}  O^{b}  (\hat{p}^{3})^{k-i} e^{i \bm{\Delta} \cdot \bm{\hat{X}}}  (\hat{p}^{3})^{i}| n \rangle.  \label{eq:moment_1b}
\end{align}
\end{subequations}
where the indices $k$ and $i$ denote powers of the momentum operator $\hat{p}^3$ in the single-particle operator, and the definition of the single-particle operator $O^{b}$ is given in Eq.~\eqref{eq:mfgpds_2}. For brevity, the indices denoting the multipole order are suppressed in Eqs.~\eqref{eq:moment_0}, \eqref{eq:moment}, and \eqref{eq:moment_1} until they become necessary.

To derive Eq.~\eqref{eq:moment} from Eq.~\eqref{eq:moment_0}, we apply integration by parts and discard the boundary contributions. This procedure transforms the weight factor $x^{m-1}$ in the integrand of Eq.~\eqref{eq:moment_0} into derivatives acting on $Z_{\mathrm{mf}}$. The distributive property of differentiation leads to the appearance of binomial coefficients in Eq.~\eqref{eq:moment}, defined as
\begin{align}
\binom{m}{k} = \frac{m!}{k!(m-k)!}, \quad \text{for} \ 0 \leq k \leq m.
\end{align}
It is important to note that, in the large-$N_{c}$ limit, when considering the partonic structure in the domain $x = O(N_c^{-1})$, the range of $x$ is not limited to $[-1, 1]$ but is extended to $x=[-\infty, \infty]$; see Refs.~\cite{Diakonov:1996sr, Diakonov:1997vc}. As a result, the boundary contributions vanish at infinity, rather than at $x = \pm 1$. For further details on the derivation of Eq.~\eqref{eq:moment}, we refer the reader to Refs.~\cite{Schweitzer:2002nm, Schweitzer:2003ms, Ossmann:2004bp, Goeke:2007fp}.

\subsection{Symmetries \label{sec:sym}}
In general, the polynomiality of GPDs is dictated by the constraints imposed by discrete symmetries—time reversal, Hermiticity, and parity—on the matrix element~\eqref{eq:General_ME_GPDs}. Likewise, the mean-field approach respects these symmetries at the level of the single-particle representation. In this section, we examine how these discrete symmetries manifest in the mean-field framework and provide a proof of polynomiality for the chiral-odd GPDs using the mean-field GPD basis.

{\textit{Time reversal ($G_{5}$ symmetry).}}
In the mean-field framework, a combination of ordinary time-reversal symmetry and an isospin rotation is referred to as the $G_{5}$ transformation~\cite{Schweitzer:2003ms}. This transformation is implemented by the unitary matrix
\begin{align}
G_{5}= \gamma^{1}\gamma^{3}\tau^{2}.
\end{align}
The Dirac and isospin matrices, as well as the single-particle Hamiltonian and wave functions, obey the following transformation rules:
\begin{subequations}
\label{eq:G5}
\begin{align}
 G_{5} \gamma^{\mu} G^{-1}_{5} &= (\gamma^{\mu})^{T}, \quad G_{5} \tau^{a} G^{-1}_{5} = -(\tau^{a})^{T},
\\[.5ex]
G_{5} \hat{H} G^{-1}_{5} &= (H)^{T}, \quad G_{5} \Phi_{n}(\bm{X}) = \Phi^{*}_{n}(\bm{X}).
\end{align}
\end{subequations}
Using these relations, along with the identity $\hat{\bm{p}} = -(\hat{\bm{p}})^{T}$, we can analyze the symmetry properties of the matrix elements involving the single-particle operators defined in Eq.~\eqref{eq:moment_1}. For a general non-diagonal matrix element, we obtain the relation
\begin{align}
& \langle j |  \Gamma (\hat{p}^{3})^{l} F(\bm{\hat{X}}) (\hat{p}^{3})^{m} | n  \rangle
\nonumber \\[.5ex]
& = (-1)^{l+m} \langle n |  (G_{5} \Gamma  G^{-1}_{5})^{T} (\hat{p}^{3})^{m} F(\bm{\hat{X}}) 
(\hat{p}^{3})^{l} | j  \rangle,
\label{eq:Gparity}
\end{align}
where $\Gamma$ is a generic spin-flavor matrix and $F(\bm{\hat{X}})$ is an arbitrary function of the displacement operator.

{\textit{Parity ($\hat{\Pi}$ symmetry).}}
Parity symmetry constrains the structure of chiral-odd GPDs by restricting the allowed angular momentum components in the partial-wave expansion of single-particle operators. The relativistic parity transformation operator $\hat{\Pi}$ is defined as
\begin{align}
&\hat{\Pi} = \hat{\Pi}^{-1} \equiv \gamma^{0}\hat{\mathcal{P}}, \quad
\hat{\mathcal{P}} F(\hat{\bm{X}}) \hat{\mathcal{P}}^{-1} = F(-\hat{\bm{X}}),
\end{align}
where $\hat{\mathcal{P}}$ denotes the spatial inversion operator acting on position-dependent functions. As discussed in Eq.~\eqref{eq:basis_a}, the single-particle Hamiltonian commutes with the parity operator, implying that the eigenstates introduced in Eq.~\eqref{eq:newf} are also eigenstates of parity:
\begin{align}
\hat{\Pi} | n \rangle = \mathcal{P}_{n} | n \rangle,
\end{align}
with eigenvalues $\mathcal{P}_{n} = \pm 1$. Under parity, the general matrix element in Eq.~\eqref{eq:Gparity} transforms as
\begin{align}
&\langle j |  \Gamma (\hat{p}^{3})^{l} F(\hat{\bm{X}}) (\hat{p}^{3})^{m} | n  \rangle = (-1)^{l+m} \mathcal{P}_{n} \mathcal{P}_{j} 
\nonumber \\[.5ex]
&\times \langle j |  (\gamma^{0}
\Gamma   \gamma^{0}) (\hat{p}^{3})^{l} (\hat{\mathcal{P}} F(\hat{\bm{X}})
\hat{\mathcal{P}}^{-1}) (\hat{p}^{3})^{m} | n  \rangle.
\label{eq:parity}
\end{align}

{\it Partial wave expansion.} 
The dependence of the moments of the mean-field GPDs in Eq.~\eqref{eq:moment} on the momentum transfer $\bm{\Delta}$ is encoded in the operator function $e^{i \bm{\Delta} \cdot \hat{\bm{X}}}$, which arises from the quantization of the translational motion of the mean field. The variable $\xi$ enters through its identification with $\Delta^3$ in the large-$N_c$ kinematics, as stated in Eq.~\eqref{eq:skew}. To make this dependence explicit, we perform a partial-wave expansion of the operator function $e^{i \bm{\Delta} \cdot \hat{\bm{X}}}$ (see Refs.~\cite{Landau:1991wop,Schweitzer:2003ms} and Appendix F of Ref.~\cite{Goeke:2007fp}):
\begin{align}
e^{i \bm{\Delta} \cdot \hat{\bm{X}}}  &= \sum^{\infty}_{l} i^{l} (2l+1) j_{l}(|\hat{\bm{X}}|\sqrt{-t}) \cr
&\times P_{l}\left(-\frac{2 \xi M_{N}}{\sqrt{-t}}\right) P_{l}\left(\frac{\hat{X}^{3}}{|\hat{\bm{X}}|}\right),
\label{eq:pwe}
\end{align}
Here, $j_{l}$ denotes the spherical Bessel functions, and $P_{l}$ are the Legendre polynomials. In particular, when taking the limit $t \to 0$ while keeping $\xi \neq 0$ fixed, the expansion reduces to
\begin{align}
\lim_{t\to 0, \xi \neq 0} e^{i \bm{\Delta}\cdot \hat{\bm{X}}} = \sum^{\infty}_{l=0} \frac{(-2i M_{N} \xi |\hat{\bm{X}}|)^{l}}{l!} P_{l} (\cos{\hat{\theta}}),
\label{eq:pwe_fw}
\end{align}
where  $\cos{\hat{\theta}} =\hat{X}^{3}/|\hat{\bm{X}}|$.

{\textit{Grand spin selection rule.}}
The mean-field GPDs in Eq.~\eqref{eq:mfgpds_1} and their moments in Eq.~\eqref{eq:moment_1} involve single and double summations over single-particle states.

The single-sum contribution, corresponding to Eq.~\eqref{eq:moment_1a}, is given by
\begin{align}
\sum_{\textrm{other}} \sum_{G,G_3}  &\langle G,G_{3} | \hat{O}_{\mu b} | G,G_{3} \rangle,
\label{eq:single_sum}
\end{align}
where ``other” denotes the sum over the remaining quantum numbers, such as the radial quantum number. Here, the bra and ket states share the same grand spin quantum numbers $G$ and $G_3$, which are summed over. In this setup, a selection rule emerges for the matrix elements of the single-particle operator $\hat{O}_{\mu b}$, which is treated as a spherical tensor operator with rank $\mu$ and projection $b$ in grand spin space. Using the Wigner–Eckart theorem, the sum over $G_3$ in Eq.~\eqref{eq:single_sum} can be evaluated as
\begin{align}
&\sum_{G,G_3}\langle G,G_{3} | \hat{O}_{\mu b} | G,G_{3} \rangle
\nonumber \\
&=  \sum_{G,G_3} (-1)^{2\mu} \frac{ \langle G G_{3}, \mu b | G G_{3}\rangle}{\sqrt{2G+1}}
\langle G || \hat{O}_{\mu} || G \rangle
\nonumber \\[1ex]
&=  \sum_{G}  \sqrt{2G+1}  \langle G || \hat{O}_{\mu} || G \rangle \delta_{\mu0} \delta_{b 0},
\label{eq:Wigner_Ekart_1}
\end{align}
where $\langle G || \hat{O}_\mu || G \rangle$ denotes the reduced matrix element. This result shows that only the scalar component of the operator—i.e., with $\mu = 0$ and $b = 0$—contributes to the matrix element. This is a  consequence of the generalized rotational invariance of the mean field.

The double-sum contribution in Eq.~\eqref{eq:moment_1b} can be expressed as
\begin{align}
\sum_{\textrm{other}}  \sum_{G,G_3,G',G'_{3}}  &\langle G,G_{3} | \tau | G',G'_{3} \rangle \cr
&\hspace{-0.4cm}\times  \langle G',G'_{3} | \hat{O} | G,G_{3} \rangle.
\label{eq:double_sum}
\end{align}
In this expression, the initial state undergoes a transition to a different state through the action of the isospin operator $\tau$, inducing a grand spin change $\Delta G = 1$. Subsequently, the system returns to the original state via the action of the single-particle operator $\hat{O}$. This process imposes specific selection rules on the matrix elements appearing in the ``double-sum" structure. For the case where $\hat{O}_{\mu b}$ is a spherical tensor operator in grand spin space, the summation over $G_3$ and $G'_3$ in Eq.~\eqref{eq:double_sum} can be evaluated using the Wigner–Eckart theorem:
\begin{align}
& \sum_{G,G_3, G',G'_{3}} \langle G,G_{3} | \tau_{1 a} | G',G'_{3} \rangle \langle G',G'_{3} | \hat{O}_{\mu b} | G,G_{3} \rangle
\nonumber \\
&=  \sum_{G,G_3, G',G'_{3}} (-1)^{2\mu} \frac{ \langle G G_{3}, 1 a | G' G'_{3}\rangle}{\sqrt{2G'+1}} \frac{ \langle G' G'_{3}, \mu b | G G_{3}\rangle}{\sqrt{2G+1}} \cr
& \times \langle G || \tau_{1} || G' \rangle\langle G' || \hat{O}_{\mu} || G \rangle
\nonumber \\[1ex]
&= \frac{1}{3}  \sum_{G, G'}  (-1)^{G'-G-a}  \langle G || \tau_{1} || G' \rangle \langle G' || \hat{O}_{\mu} || G \rangle \delta_{1 \mu} \delta_{a-b},
\label{eq:Wigner_Ekart_2}
\end{align}
where the identity for the contraction of two Clebsch–Gordan coefficients is applied:
\begin{align}
&\sum_{G'_{3}G_{3}} \langle G G_{3}, 1 a | G' G'_{3}\rangle \langle G' G'_{3}, \mu b | G G_{3}\rangle \cr
&= (-1)^{G-G'-a} \frac{\sqrt{2G'+1}\sqrt{2G+1}}{{2 \mu+1}} \delta_{1 \mu} \delta_{a -b}.
\end{align}
This result shows that only the tensor component with rank $\mu = 1$ and projection $a + b = 0$ contributes to the matrix element. 

\subsection{Polynomiality}
Making use of the symmetries discussed in Sec.~\ref{sec:sym}, we now demonstrate the polynomiality property of the chiral-odd GPDs within the basis of the mean-field GPDs. For the monopole mean-field GPD, the single- and double-sum matrix elements in Eq.~\eqref{eq:moment_1} of the single-particle operator take the form:
\begin{subequations}
\label{eq:poly_1}
\begin{align}
A_{0}^{ki}(\xi,t) &= i \epsilon^{3ab} \langle n| \tau^{b} | j \rangle \langle j | (1+\gamma^{0} \gamma^{3}) \gamma^{a} \cr
&\times(\hat{p}^{3})^{k-i} e^{i \bm{\Delta} \cdot \hat{\bm{X}}} (\hat{p}^{3})^{i} | n \rangle, \\[1ex]
B_{0}^{ki}(\xi,t) &= i \epsilon^{3ab} \langle n | \tau^{b} (1+\gamma^{0} \gamma^{3}) \gamma^{a} \cr
&\times (\hat{p}^{3})^{k-i} e^{i \bm{\Delta} \cdot \hat{\bm{X}}} (\hat{p}^{3})^{i} | n \rangle.
\end{align}
\end{subequations}
Using the $G_5$-symmetry~\eqref{eq:Gparity}, the matrix elements transform as:
\begin{subequations}
\label{eq:poly_2}
\begin{align}
A_{0}^{ki}(\xi,t) &= (-1)^{k+1} i \epsilon^{3ab} \langle n| \tau^{b} | j \rangle  \langle j | (1-\gamma^{0} \gamma^{3}) \gamma^{a} \cr
 &\times (\hat{p}^{3})^{k-i} e^{i \bm{\Delta} \cdot \hat{\bm{X}}} (\hat{p}^{3})^{i} | n \rangle, \\[1ex]
B_{0}^{ki}(\xi,t) &= (-1)^{k+1} i \epsilon^{3ab} \langle n | \tau^{b} (1-\gamma^{0} \gamma^{3}) \gamma^{a} \cr
&\times (\hat{p}^{3})^{k-i} e^{i \bm{\Delta} \cdot \hat{\bm{X}}} (\hat{p}^{3})^{i} | n \rangle.
\end{align}
\end{subequations}
By comparing Eqs.~\eqref{eq:poly_1} and \eqref{eq:poly_2}, we observe that the contribution from the identity matrix $1$ or the Dirac matrix $\gamma^{0} \gamma^{3}$ in the combination $(1 + \gamma^{0} \gamma^{3})$ depends on the parity of the index $k$. Specifically, only the term $1$ survives for odd $k$, and $\gamma^{0} \gamma^{3}$ survives for even $k$. As a result, the matrix elements simplify to:
\begin{subequations}
\label{eq:poly_3}
\begin{align}
&A_{0}^{ki}(\xi,t) \cr
&=  i \epsilon^{3ab} \langle n| \tau^{b} | j \rangle \langle j | (\gamma^{0} \gamma^{3})^{k+1} \gamma^{a} e^{i \bm{\Delta} \cdot \hat{\bm{X}}} (\hat{p}^{3})^{i} | n \rangle, \\
&B_{0}^{ki}(\xi,t)  \cr
&=  i \epsilon^{3ab} \langle n | \tau^{b} (\gamma^{0} \gamma^{3})^{k+1} \gamma^{a} e^{i \bm{\Delta} \cdot \hat{\bm{X}}} (\hat{p}^{3})^{i} | n \rangle.
\end{align}
\end{subequations}

Next, we examine the effect of the parity transformation. To this end, we first perform a partial-wave expansion. Applying the identity~\eqref{eq:pwe} to Eq.~\eqref{eq:poly_2}, we obtain:
\begin{subequations}
\label{eq:poly_4}
\begin{align}
A_{0}^{ki}(\xi,t) &=   \sum^{\infty}_{l=0} i^{l} (2l+1)
P_{l}\left(-\frac{2 \xi M_{N}}{\sqrt{-t}}\right) \cr
&\times  i \epsilon^{3ab} \langle n| \tau^{b} | j \rangle \langle j | (\gamma^{0} \gamma^{3})^{k+1} \gamma^{a} (\hat{p}^{3})^{k-i} \cr
&\times  j_{l}(|\hat{\bm{X}}|\sqrt{-t})
P_{l}(\cos{\hat{\theta}}) (\hat{p}^{3})^{i} | n \rangle,  \label{eq:poly_4a}\\[1ex]
B_{0}^{ki}(\xi,t) &=   \sum^{\infty}_{l=0} i^{l} (2l+1)
P_{l}\left(-\frac{2 \xi M_{N}}{\sqrt{-t}}\right) \cr
&\times i \epsilon^{3ab} \langle n | \tau^{b} (\gamma^{0} \gamma^{3})^{k+1} \gamma^{a} (\hat{p}^{3})^{k-i} \cr
&\times  j_{l}(|\hat{\bm{X}}|\sqrt{-t})
P_{l}(\cos{\hat{\theta}}) (\hat{p}^{3})^{i} | n \rangle. \label{eq:poly_4b}
\end{align}
\end{subequations}
Now, applying the identity~\eqref{eq:parity} for the parity transformation to Eq.~\eqref{eq:poly_4}, we find that the matrix elements vanish for odd values of $l$. This implies that the moment of the monopole mean-field GPD forms a polynomial in an infinite series of even powers of $\xi$, leading to a modification of the sum over $l$ as
\begin{align}
\sum^{\infty}_{l=0} [\ldots ] \to \sum^{\infty}_{l=0,2,4,\ldots} [\ldots ].
\end{align}
Here, we encounter an infinite series in $\xi$ due to the unbounded extent of $l$. However, the grand spin selection rules discussed in Section~\ref{sec:sym} impose an upper limit on the allowed values of $l$. To see this, we must first examine the rank of the single-particle operator. 

For the spin-isospin part of the single-particle operator in Eq.~\eqref{eq:poly_4a}, we simplify the 2D operator representation to its three-dimensional (3D) form in order to more easily identify its rank in the grand spin basis:
\begin{subequations}
\label{eq:gamma_i}
\begin{align}
i\epsilon^{3ab} (\gamma^{0}\gamma^{3})^{k+1} \gamma^{a} &= \gamma^{0} \Sigma^{b}_{\perp}&& (k \ \text{even}),  \\[.5ex]
&= i\epsilon^{3ab} \gamma^{a} && (k \ \text{odd}).
\end{align}
\end{subequations}
From Eq.~\eqref{eq:gamma_i}, we find that the maximal rank of the spin-isospin operator is rank-1. We then couple this operator to the position-space operator (orbital angular momentum) in Eq.~\eqref{eq:poly_4}. The maximal rank of the orbital angular momentum operator is $k + l $ for both even and odd $k$. Combining the maximal rank of the spin-isospin operator in Eq.~\eqref{eq:gamma_i} with that of the orbital angular momentum part, we determine that the maximal rank of the single-particle operator in Eq.~\eqref{eq:poly_4a} is $k + l + 1$ in the grand spin quantum number. Making use of the selection rule~\eqref{eq:Wigner_Ekart_2} for the ``double sum,'' this maximal rank must be equal to rank-$1$. This condition determines the maximal value of $l_{\mathrm{max}}$. The same analysis can also be applied to the single-sum matrix element~\eqref{eq:poly_4b} (cf.~\cite{Kim:2024ibz}), and we finally derive the upper bounds of the partial waves, $l^{A}_{\mathrm{max}}$ and $l^{B}_{\mathrm{max}}$, for Eqs.~\eqref{eq:poly_4a} and~\eqref{eq:poly_4b}, respectively:
\begin{subequations}
\label{l_max_m_2_main}
\begin{alignat}{2}
l^{A}_{\mathrm{max}}(k) &= k+2, \\[0.5ex]
l^{B}_{\mathrm{max}}(k) &= k+2 \quad (k \ \text{even}), \\[0.5ex] 
&= k+1 \quad (k \ \text{odd}).  
\end{alignat}
\end{subequations}
In the expression for the $m$-th moment given in Eq.~\eqref{eq:moment}, the summation over $k$ runs from $0$ to $m - 1$ for Eq.~\eqref{eq:poly_4a}, and from $0$ to $m - 2$ for Eq.~\eqref{eq:poly_4b}. For a fixed $m$, the highest partial wave $l$ that appears in the expansions of Eq.~\eqref{eq:poly_4a} and Eq.~\eqref{eq:poly_4b} corresponds to the cases $k = m - 1$ and $k = m - 2$, respectively. Based on Eq.~\eqref{l_max_m_2_main}, these maximal partial waves are given by
\begin{alignat}{2}
\label{l_max_m_3_main}
l^{A}_{\mathrm{max}}(m) &= m+1,
\end{alignat}
for Eq.~\eqref{eq:poly_4a}, and
\begin{subequations}
\label{l_max_m_4}
\begin{alignat}{2}
l^{B}_{\mathrm{max}}(m) &= m \quad &&(m \ \text{even}), \\[0.5ex] 
&= m-1 \quad &&(m \ \text{odd}),  
\end{alignat}
\end{subequations}
for Eq.~\eqref{eq:poly_4b}. Note that the nonvanishing contribution to $B_{0}$ starts from $m \geq 2$ due to the overall factor of $m - 1$ in Eq.~\eqref{eq:moment}. 

Taking into account the highest partial waves determined above, we finally arrive at the results dictated by the discrete symmetries:
\begin{subequations}
\label{eq:poly_5}
\begin{align}
A_{0}^{ki}(\xi,t) &=   \sum^{l^{A}_{\mathrm{max}}(m)}_{l=\mathrm{even}} i^{l} (2l+1)
P_{l}\left(-\frac{2 \xi M_{N}}{\sqrt{-t}}\right)  \cr
&\times i \epsilon^{3ab} \langle n| \tau^{b} | j \rangle  \langle j | (\gamma^{0} \gamma^{3})^{k+1} \gamma^{a} \nonumber \\[1ex]
&\times (\hat{p}^{3})^{k-i} j_{l}(|\hat{\bm{X}}|\sqrt{-t})
P_{l}(\cos{\hat{\theta}}) (\hat{p}^{3})^{i} | n \rangle, \\
B_{0}^{ki}(\xi,t) &=   \sum^{l^{B}_{\mathrm{max}}(m)}_{l=\mathrm{even}} i^{l} (2l+1)
P_{l}\left(-\frac{2 \xi M_{N}}{\sqrt{-t}}\right) \cr
&\times i \epsilon^{3ab} \langle n | \tau^{b} (\gamma^{0} \gamma^{3})^{k+1} \gamma^{a} \nonumber \\[1ex]
&\times(\hat{p}^{3})^{k-i} j_{l}(|\hat{\bm{X}}|\sqrt{-t})
P_{l}(\cos{\hat{\theta}}) (\hat{p}^{3})^{i} | n \rangle.
\end{align}
\end{subequations}
We observe that the polynomial properties of the moments are governed by the Legendre polynomials associated with the angular momentum $ l $. The moments of the GPDs become even functions of $ \xi $, and the maximal power of $ \xi $ in the polynomial is $ m + 1 $. This is two powers higher than the $ m - 1 $ bound in Eq.~\eqref{eq:polynomiality}, reflecting the fact that the monopole mean-field GPD includes an additional $ \xi^{2} $ dependence, i.e., $ Z_{\mathrm{mf},0} \propto \xi^{2} E_{T} $. We have thus proven the polynomiality of the monopole mean-field GPD. The polynomiality of higher-multipole mean-field GPD is discussed in Appendix~\ref{app:poly}.

\subsection{Sum rules}
We now establish a direct correspondence between the moments of the mean-field GPDs and the tensor form factors, which is consistent with the standard sum rules~\eqref{eq:first_Mel}. Explicit expressions for the mean-field tensor form factors and their relations to the standard tensor form factors are presented in Appendix~\ref{app:tff}.

The first moment of the monopole GPD is given by Eq.~\eqref{eq:poly_5} with $m = 1$,
\begin{align}
\int dx \, Z_{\mathrm{mf},0}(x, \xi, t) = - \frac{N_{c}}{4I} \sum_{\substack{n,\mathrm{non} \\ j,\mathrm{occ}}} \frac{1}{E_{n}-E_{j}} A^{00}_{0}(\xi,t),
\label{eq:sr_1}
\end{align}
where the single-sum contribution vanishes. The double-sum contribution $A^{00}_{0}(\xi,t)$, in which the allowed partial waves run over $l = 0, 2$, is derived as
\begin{align}
A^{00}_{0}(\xi,t) &= \langle n | \bm{\tau_{\perp}} | j \rangle \cdot \langle j | \gamma^{0}\bm{\Sigma_{\perp}} j_{0}(|\hat{\bm{X}}| \sqrt{-t}) | n \rangle \nonumber \\[1ex]
&+ \left(\frac{15}{4} + \frac{45 M^{2}_{N} \xi^{2}}{t}\right) \langle n | \bm{\tau_{\perp}} | j \rangle \cdot \langle j | \gamma^{0} \bm{\Sigma_{\perp}} \nonumber \\[1ex]
&\times Y^{33}_{2}(\Omega_{\hat{\bm{X}}}) j_{2}(|\hat{\bm{X}}| \sqrt{-t})  | n \rangle,
\label{eq:sr_2_main}
\end{align}
by using the relations for the Dirac matrices~\eqref{eq:gamma_i} and the Legendre polynomial of degree $l = 2$, i.e., $P_{2}(x) = (3x^{2} - 1)/2$. Here, $Y_{2}$ denotes the 3D irreducible rank-2 tensor, defined as
\begin{align}
Y^{ab}_{2} (\Omega_{\bm{\hat{X}}})  \equiv \frac{{\hat{X}}^{a}{\hat{X}}^{b}}{|\bm{\hat{X}}|^{2}} - \frac{1}{3}\delta^{ab}.
\label{eq:3dto2d}
\end{align}
The results in Eq.~\eqref{eq:sr_2_main} are expressed in a mixed representation involving both 2D and 3D irreducible tensors. However, the grand spin selection rule governing the double sum allows us to reexpress them in a 3D rotationally invariant form. Using Eq.~\eqref{eq:Wigner_Ekart_2}, the first and second terms in Eq.~\eqref{eq:sr_2_main} can respectively be rewritten as
\begin{align}
&\langle n | \bm{\tau}_{\perp} | j \rangle \cdot \langle j | \gamma^{0} \bm{\Sigma}_{\perp} j_{0}(|\hat{\bm{X}}|\sqrt{-t}) | n \rangle \cr
&= \frac{2}{3}\langle n | \bm{\tau} | j \rangle \cdot \langle j | \gamma^{0} \bm{\Sigma} j_{0}(|\hat{\bm{X}}|\sqrt{-t}) | n \rangle, 
\end{align}
and
\begin{align}
&\langle n | \bm{\tau}_{\perp} | j \rangle \cdot \langle j | \gamma^{0} \bm{\Sigma}_{\perp} Y^{33}_{2}(\Omega_{\hat{\bm{X}}}) j_{2}(|\hat{\bm{X}}|\sqrt{-t}) | n \rangle \cr
&= -\frac{2}{15}\langle n | {\tau}^{a} | j \rangle  \langle j | \gamma^{0} {\Sigma}^{b} Y^{ab}_{2}(\Omega_{\hat{\bm{X}}}) j_{2}(|\hat{\bm{X}}|\sqrt{-t}) | n \rangle.
\end{align}
With these relations, we can express the first moment of the monopole GPD in terms of the mean-field form factors~\eqref{eq:local} derived from the local operator. This leads to
\begin{align}
&\int dx \, Z_{\mathrm{mf},0}(x, \xi, t) \cr
&=N_{\mathrm{mf},0} - \left(\frac{t}{24M^{2}_{N}} + \frac{\xi^{2}}{2}\right) N_{\mathrm{mf},2}.
\end{align}
Using the relation~\eqref{eq:local_1}, we replace the mean-field form factors with the standard tensor form factors:
\begin{align}
&\int dx \, Z_{\mathrm{mf},0}(x, \xi, t) \cr
&= H^{u+d}_{T} + \left(\frac{t}{8M^{2}_{N}} - \frac{\xi^{2}}{2} \right)E^{u+d}_{T}.
\end{align}
This expression matches the first moment of the monopole GPD defined in Eq.~\eqref{relation_0}, where the first moment of $\tilde{E}_{T}$ vanishes. We thus confirm the validity of the sum rule within the mean-field framework. A detailed derivation of the sum rules for the higher-multipole mean-field GPDs is presented in Appendix~\ref{app:sr}.

\section{Gradient expansion \label{sec:ge}}
In the previous section, we demonstrated that all general constraints—such as polynomiality and sum rules—on the chiral-odd GPDs are satisfied within the mean-field framework. This enhances the reliability of the mean-field picture for making numerical predictions. In this context, we are now prepared to perform numerical estimations of the mean-field chiral-odd GPDs.

Although a complete prediction requires summing over all single-particle wave functions, we instead employ the gradient expansion to simplify the computation and gain a clearer physical understanding for the Dirac sea contributions~(or chiral dynamics). Rather than expressing the quark propagator in terms of a single-particle basis for the disconnected diagram (see Fig.~\ref{fig:dia}), we expand it in powers of derivatives, $\partial U \ll M$, in what is known as the chiral expansion. We then retain the non-vanishing leading contributions to the matrix elements. This technique is also used in deriving the fully bosonized effective chiral action, where both the Gasser–Leutwyler terms with four derivatives and the Wess–Zumino–Witten terms with correct coefficients naturally emerge. This approach yields analytic expressions for the matrix elements in terms of the chiral (pion) field and enables the investigation of dominant chiral contributions. As such, it serves as a valuable complement to the predictions of chiral perturbation theory.

Maintaining the representation of the quark propagator~\eqref{eq:action}, we derive the Dirac sea contribution to the NLO matrix element of the effective chiral-odd operator~\eqref{eq:eff} between baryon states in the $1/N_c$ expansion. The LO matrix element was derived in Refs.~\cite{inprep}, and here we focus solely on the NLO contributions. The spin-flavor structure discussed in the single-particle representation in Sec.~\ref{sec:sfs} is clearly preserved in the current representation. Focusing on proton states, we obtain
\begin{align}
\left.\begin{array}{c}  \mathcal{M}^{u+d}[i\sigma^{+j}] \\[2ex] \mathcal{M}^{u-d}[i\sigma^{+j}] \end{array} \right\}
&\overset{\mathrm{NLO}}{=} \frac{M^{2}_{N}  N_{c}}{2\sqrt{2}I M}  \left\{\begin{array}{c}  2J^{k}_{S'_3 S_3} \\[2ex] -\frac{1}{3} \delta_{S'_3 S_3}  \end{array} \right\} \nonumber  \\[0.5ex]
&\times \int \frac{d^{4}p}{(2\pi)^{4}}  \left\{\begin{array}{c}  \Pi^{u+d,jk} \\[2ex] \Pi^{u-d,j} \end{array} \right\} ,
\label{eq:ge_1}
\end{align}
where $\Pi \equiv \Pi(x,\xi,p,\bm{\Delta})$ is a correlator transformed into momentum space in the background of the mean field, and is defined as
\begin{align}
&\left\{\begin{array}{c}  \Pi^{u+d,jk} \\[2ex] \Pi^{u-d,j} \end{array} \right\}  \cr
&=i M \langle \bm{p} - \bm{\Delta} | \bigg{\{} \, \delta(M_{N}(x-\xi) -p\cdot v) \cr& \times \mathrm{tr}\bigg{[}(1+\gamma^{0}\gamma^{3} )\gamma^{j}   \left\{\begin{array}{c} \bm{1} \\[2ex] \tau^{k} \end{array} \right\} \frac{1}{p^{0}- H} \tau^{k} \frac{1}{p^{0}- H}  \bigg{]} \cr
&+\frac{d}{d x M_{N}} \, \delta(M_{N}(x-\xi) -p\cdot v) \cr
&\times \mathrm{tr}\bigg{[}(1+\gamma^{0}\gamma^{3} )\gamma^{j}   \left\{\begin{array}{c} \bm{1} \\[2ex] \tau^{k} \end{array} \right\} \frac{1}{p^{0}- H} \tau^{k}  \bigg{]} \bigg{\}} | \bm{p} \rangle, 
\label{eq:ge_2}
\end{align}
with $v^{\mu} = (1,0,0,-1)$ the light-cone vector. The trace is taken over both flavor and Dirac spaces. After performing zero-mode quantization and considering the contribution linear in the angular velocity, the resulting expression involves two quark propagators with a flavor matrix inserted between them.

Having projected Eq.~\eqref{eq:ge_1} onto the multipole structures in Eq.~\eqref{eq:model_spin_flavor_2}, we derive the mean-field GPDs as
\begin{align}
&Z_{\mathrm{mf}} = - \frac{M_{N}  N_{c}}{8I M}   \int \frac{d^{4}p}{(2\pi)^{4}}  \mathcal{C}, 
\label{eq:ge_3}
\end{align}
with 
\begin{align}
\mathcal{C}&= i\epsilon^{3jk} \Pi^{u+d,jk} && \left(Z_{\mathrm{mf},0}\right), \cr
&= \frac{4}{3} \frac{\Delta^{j}_{\perp}}{|\bm{\Delta}_{\perp}|^{2}} \Pi^{u-d,j} && \left(\frac{Z_{\mathrm{mf},1}}{M_{N}}\right), \cr
&= 4i\epsilon^{3jl} \delta^{3k} \frac{\Delta^{l}_{\perp}}{|\bm{\Delta}_{\perp}|^{2}} \Pi^{u+d,jk} &&\left(\frac{\tilde{Z}_{\mathrm{mf},1}}{M_{N}}\right), \cr
&= 16 i \epsilon^{3jp} \frac{1}{|\bm{\Delta}_{\perp}|^{4}}  \cr
&\times \left(\Delta^{p}_{\perp}\Delta^{k}_{\perp}- \frac{1}{2} \delta^{pk} |\bm{\Delta}_{\perp}|^{2}\right) \Pi^{u+d,jk} && \left(\frac{Z_{\mathrm{mf},2}}{M^{2}_{N}}\right).
\label{eq:ge_4}
\end{align}
As in Eqs.~\eqref{eq:mfgpds_1} and \eqref{eq:mfgpds_2}, the multipole characteristics manifest in the multipole correlator $\mathcal{C}$.

To proceed, we perform the operator expansion\footnote{ In the first line of Eq.~\eqref{eq:ge_5}, the numerator and denominator are evidently commutable and symmetric. However, after performing the gradient expansion and truncating it, the hermiticity of these operators is no longer manifest (see Ref.~\cite{Ossmann:2004bp}). To address this, we symmetrize the expression as follows:
\begin{align}
\frac{1}{p^{0}-H}= \frac{1}{2}\left\{(p^{0} + H), \frac{1}{\hat{p}^{2}-M^{2}-i M \bm{\gamma}\cdot \bm{\nabla} U^{\gamma_{5}}} \right\} ,
\end{align}
and then expand it in powers of $i M \bm{\gamma} \bm{\nabla} U^{\gamma_{5}}$. }: the quark propagators in Eq.~\eqref{eq:ge_2} are expanded in powers of the pion momentum, yielding a bosonized representation of the correlator and generating an infinite tower of pion-momentum contributions:
\begin{align}
\frac{1}{p^{0}-H} &= (p^{0} + H) \frac{1}{\hat{p}^{2}-M^{2}-i M \bm{\gamma} \cdot\bm{\nabla} U^{\gamma_{5}}} \cr
&= (p^{0} + H) \frac{1}{D(\hat{p})} \cr
& + (p^{0} + H) \frac{1}{D(\hat{p})} (i M \bm{\gamma} \cdot \bm{\nabla} U^{\gamma_{5}}) \frac{1}{D(\hat{p})} + ...,
\label{eq:ge_5}
\end{align}
where the 4-momentum operator is defined as $\hat{p} = (p^{0}, \hat{\bm{p}})$, and the free propagator $1/D(\hat{p})$ is given by
\begin{align}
\frac{1}{D(\hat{p})} = \frac{1}{\hat{p}^{2}-M^{2} + i 0},
\label{eq:ge_6}
\end{align}
with the standard $i0$ prescription. Inserting the symmetrized propagator~\eqref{eq:ge_5} into Eq.~\eqref{eq:ge_2}, we present the correlator $\Pi$ as a sum over an infinite series of bosonized correlators $\Pi_{\alpha}$ in momentum space:
\begin{align}
\Pi^{u\pm d}&= \delta(M_{N}(x-\xi) -p\cdot v)   \sum^{\infty}_{\alpha} \Pi^{u\pm d}_{\alpha} \cr
&    \pm \bigg{(}\xi \to -\xi, \bm{\Delta} \to -\bm{\Delta}\bigg{)},
\label{eq:ge_7}
\end{align}
where $\alpha$ denotes the order of derivatives acting on the chiral field ($\partial^{\alpha} U$) included in the bosonized correlators.

\subsection{Leading contributions}
\begin{figure}[t]
\includegraphics[scale=0.3]{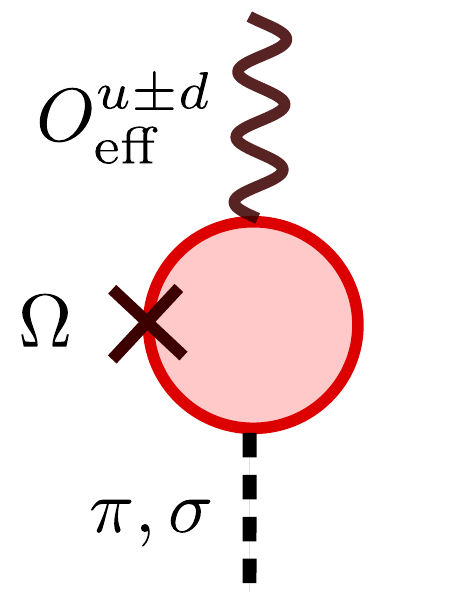} \hspace{0.5cm}
\includegraphics[scale=0.3]{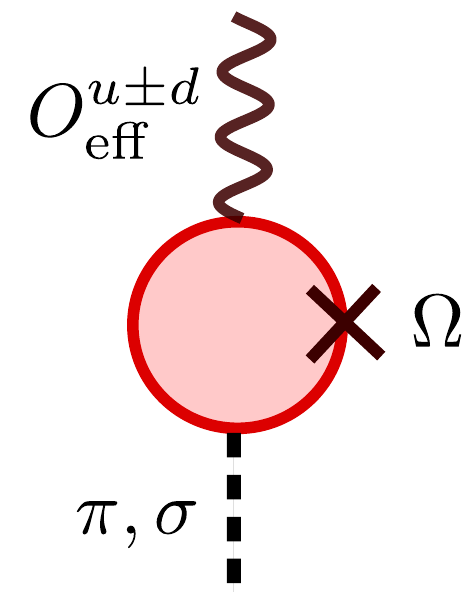}
\caption{In the NLO contributions to the chiral-odd GPDs in the $1/N_c$ expansion, the quark loop involves the effective chiral-odd operator $O^{u\pm d}_{\mathrm{eff}}$ and the angular velocity $\Omega$~\eqref{eq:av}, which induces an isovector operator under rigid rotation. This mechanism gives rise to single-pion ($\pi$) and single-scalar-meson ($\sigma$) exchange contributions to the chiral-odd GPDs.}
\label{fig:w}
\end{figure}

Having expanded the correlator~\eqref{eq:ge_7} and taken the nonvanishing leading contribution with $\alpha=0$, we obtain
\begin{align}
\Pi^{u+d,jk}_{\alpha=0} &=  \frac{M^{2} F(p-\Delta)F(p)}{2D(p-\Delta)D(p)}   \cr
& \times \epsilon^{3ji} \Delta^{i} \mathrm{tr}[\tau^{k} \{ \widetilde{U}(-\bm{\Delta}) - \widetilde{U^{\dagger}}(-\bm{\Delta}) \} ], \cr
\Pi^{u-d,j}_{\alpha=0} &= - \frac{3M^{2} F(p-\Delta)F(p)}{2D(p-\Delta)D(p)} \cr
&\times i \Delta^{j}  \mathrm{tr}[  \widetilde{U}(-\bm{\Delta}) + \widetilde{U^{\dagger}}(-\bm{\Delta}) ] - (U\to1),   
\label{eq:lc_1}
\end{align}
where the trace runs over flavor space. Here, the momentum-space representation of the chiral field is defined as the Fourier transform of its position-space counterpart and is denoted with a tilde:
\begin{align}
\widetilde{U}(\bm{k}) \equiv \int d^{3} x \, e^{-i \bm{k} \cdot \bm{x}} U(\bm{x}).
\label{eq:ldb_1}
\end{align}
The leading contribution~\eqref{eq:lc_1} originates from a single chiral field (see Fig.~\ref{fig:w}) and depends solely on the 3D momentum transfer $\bm{\Delta}$, being independent of the quark loop momentum $p$. Importantly, the quark loop integral is UV divergent, but this divergence is naturally regulated by the momentum-dependent dynamical quark mass functions $F(p)$ appearing in the numerator. The leading contribution to the mean-field GPDs in the gradient expansion is obtained by contracting the expression from Eq.~\eqref{eq:lc_1} with each of the multipole structures in Eq.~\eqref{eq:ge_4}.

\subsection{Subleading contributions \label{app:sc}} 
The UV divergent contribution arises not only from $\alpha = 0$ but also from $\alpha = 1$. For completeness of our study for the UV divergence contributions, we also examine the subleading contribution in the gradient expansion. Expanding to $\alpha = 1$, we obtain
\begin{align}
&\Pi^{u+d,jk}_{\alpha=1}  =  -\frac{M^{2} F(p-\Delta)F(p)}{D(p-\Delta)D^{2}(p)} \cr
&\times \bigg{[}  \epsilon^{3ja} (2(p^{0})^{2} + 2M^{2} + \bm{p}^{2}) +  \epsilon^{jab} (4p^{b}-\Delta^{b}) p^{0} \cr
& - (\delta^{ab} \epsilon^{3jc} - \delta^{bc} \epsilon^{3ja} + \delta^{ac} \epsilon^{3jb} ) (p^{b}-\Delta^{b}) p^{c} \bigg{]} \cr
&\times \mathrm{tr}[i \tau^{k} \{\widetilde{\partial_{a}U}(-\bm{\Delta}) - \widetilde{\partial_{a}U^{\dagger}}(-\bm{\Delta}) \} ] \nonumber \\[2ex]
&+ \mathcal{O}(\text{UV finite}),
\label{eq:sl_1}
\end{align}
for the flavor singlet, and
\begin{align}
&\Pi^{u-d,j}_{\alpha=1} =  -\frac{3M^{2} F(p-\Delta)F(p)}{D(p-\Delta)D^{2}(p)} \cr
&\times\bigg{[}  \delta^{ja} (2(p^{0})^{2} + 2M^{2} + \bm{p}^{2}) \cr
& + (\delta^{3c}\delta^{ja}-\delta^{3a}\delta^{jc}) (4p^{c}-\Delta^{c}) p^{0} \cr
& - (\delta^{jb} \delta^{ac}- \delta^{ja}\delta^{bc} + \delta^{jc}\delta^{ba} ) (p^{b}-\Delta^{b}) p^{c}   \bigg{]} \cr
&\times \mathrm{tr}[  \widetilde{\partial_{a}U}(-\bm{\Delta}) + \widetilde{\partial_{a}U^{\dagger}}(-\bm{\Delta})   ] - (U\to 1)  \nonumber \\[2ex]
&+ \mathcal{O}(\text{UV finite}),
\label{eq:sl_2}
\end{align}
for the flavor non-singlet. Unlike the case of $\alpha = 0$, the subleading contribution includes UV-finite terms, such as the mass term $M^{2}$ in Eqs.~\eqref{eq:sl_1} and \eqref{eq:sl_2}. In addition, contributions from more than one chiral field appear, all of which are UV finite. Since UV divergences arise only from terms involving a single chiral field, we retain only those contributions in this work. The subleading contribution to the mean-field GPDs in the gradient expansion is computed by contracting the expressions in Eqs.~\eqref{eq:sl_1} and \eqref{eq:sl_2} with each of the multipole structures in Eq.~\eqref{eq:ge_4}.

\section{Numerical results \label{num}}

Now, we estimate the chiral-odd GPDs within the mean-field picture. Before presenting the numerical results, we first describe the numerical setup. The self-consistent mean-field solution is well approximated by the arctangent profile function, which reproduces the correct long-distance behavior~\cite{Diakonov:1987ty, Diakonov:1988mg, Diakonov:1996sr}. Accordingly, we adopt the following form for the profile function:
\begin{align}
P(r) = -2 \arctan\left(\frac{R^{2}_{0}}{r^{2}}\right),  && MR_{0} = 1,
\label{eq:profile}
\end{align}
where $R_{0}$ is the spatial size of the mean-field and is of order $\sim M^{-1}$. The value of the dynamical quark mass is adotped to be $M = 350$~MeV, which is determined by solving the gap equation within the configuration of the instanton ensemble ($\bar{\rho} \approx 1/3$ fm, $\bar{R} \approx 1$ fm). Using these parameter sets, the nucleon mass is determined to be~\cite{Diakonov:1996sr}
\begin{align}
M_{N}=1170~\mathrm{MeV}, \quad  I\approx 1.1~\mathrm{fm}.
\label{eq:nu_mass}
\end{align}
In addition, instead of employing the PV regularization scheme, we take into account the momentum-dependent dynamical quark mass in this work. This is characterized by the quark zero-mode form factor at a low normalization point, which is predicted by the QCD instanton vacuum. This form factor is well approximated by the following simple form~\cite{Petrov:1998kg, Praszalowicz:2001wy} in Euclidean space:
\begin{align}
F(p^{2}) = \frac{-\Lambda^{2}}{p^{2} + \Lambda^{2}},  &&\text{with}  \quad \Lambda = 2^{\frac{1}{2}} \bar{\rho}^{-1}.
\label{eq:zmff}
\end{align}

\begin{figure}[t]
\includegraphics[scale=0.29]{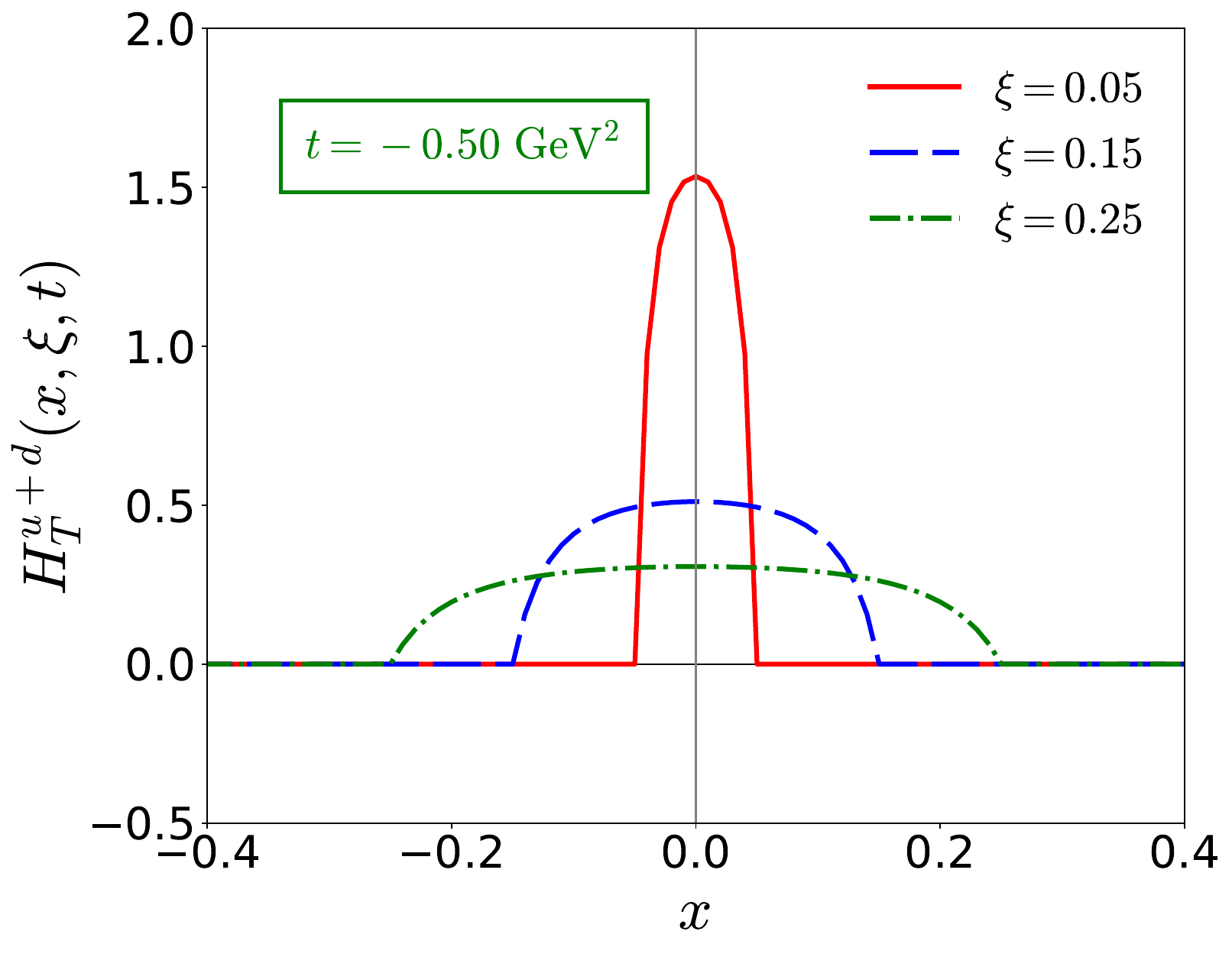} 
\includegraphics[scale=0.29]{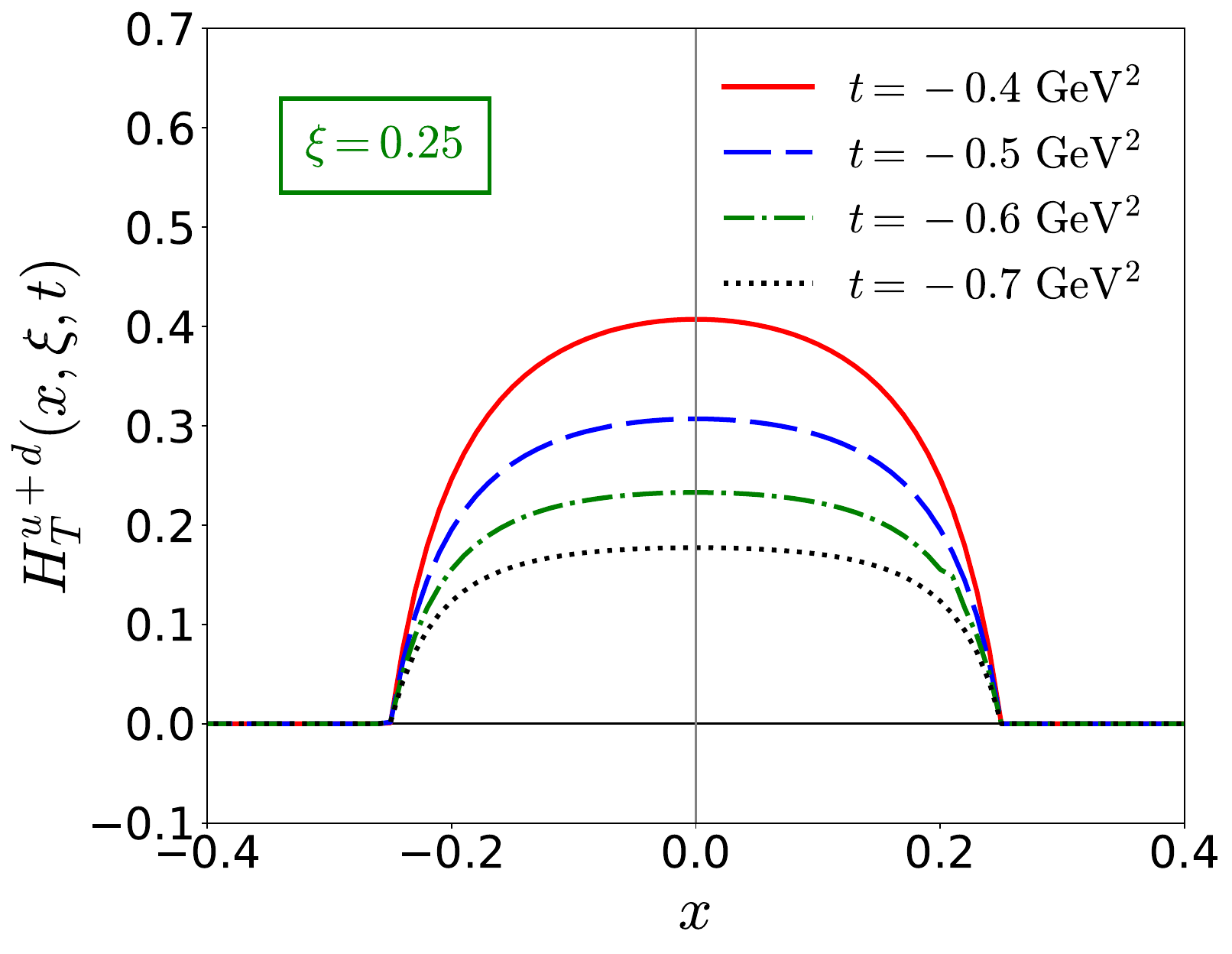}
\caption{Numerical results for the chiral-odd GPD $H^{u+d}_{T}$ at fixed $t = -0.5~\mathrm{GeV}^{2}$ with varying $\xi$ (upper panel), and at fixed $\xi = 0.25$ with varying $t$ (lower panel).}
\label{fig:3}
\end{figure}
\begin{figure}[t]
\includegraphics[scale=0.29]{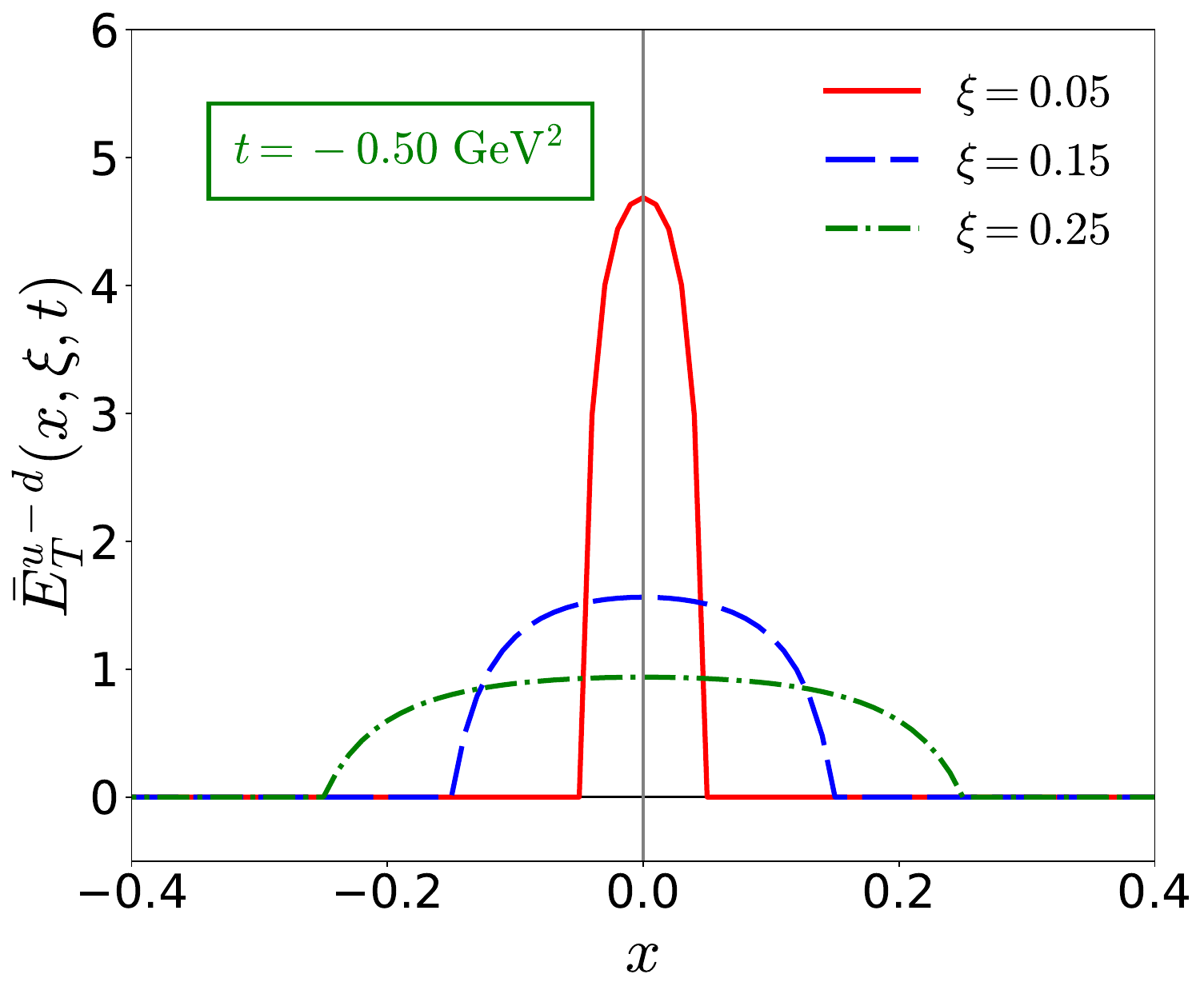} 
\includegraphics[scale=0.29]{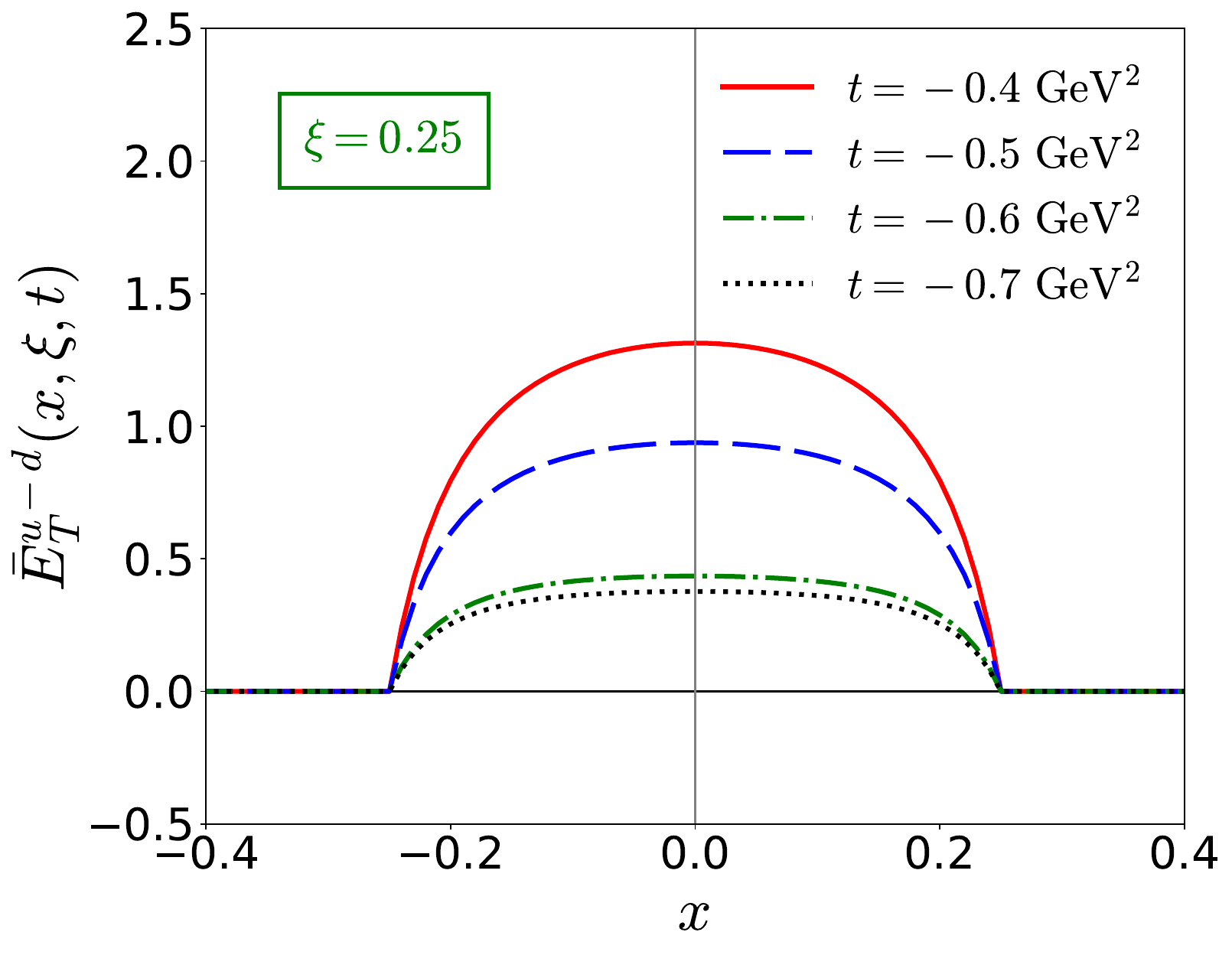}
\caption{Numerical results for the chiral-odd GPD $\bar{E}^{u-d}_{T}$ at fixed $t = -0.5~\mathrm{GeV}^{2}$ with varying $\xi$ (upper panel), and at fixed $\xi = 0.25$ with varying $t$ (lower panel).}
\label{fig:4}
\end{figure}
\begin{figure}[t]
\includegraphics[scale=0.29]{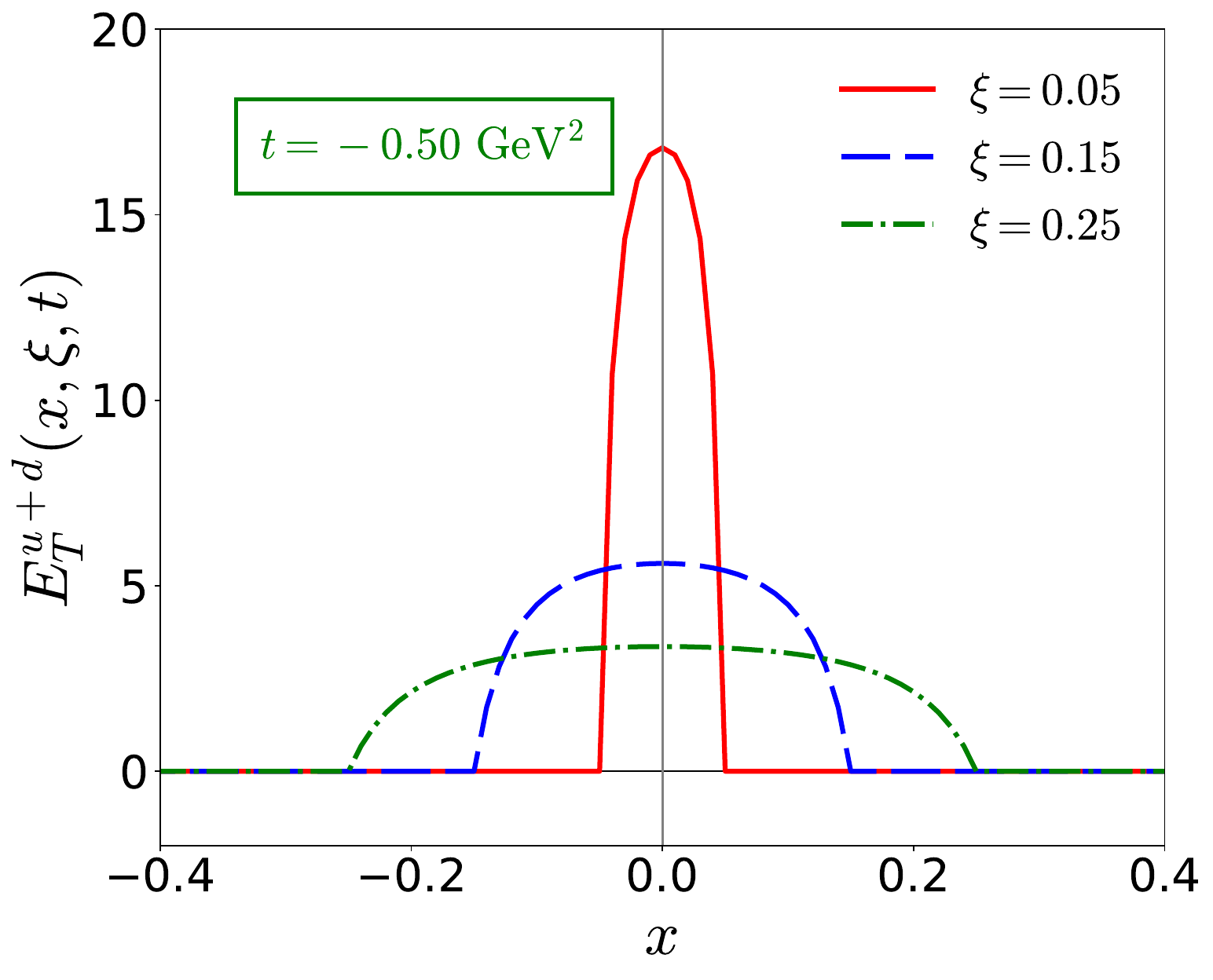} 
\includegraphics[scale=0.29]{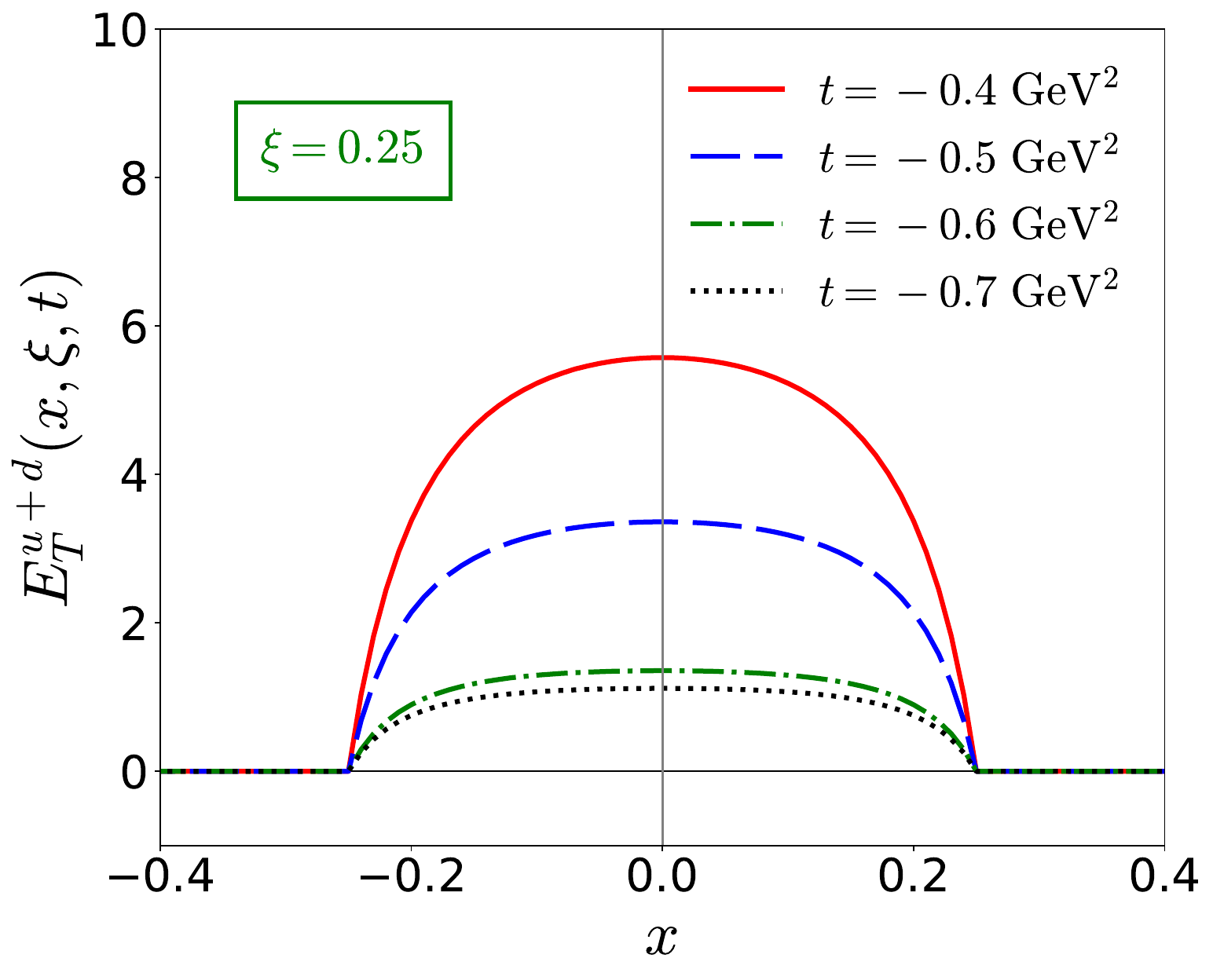}
\caption{Numerical results for the chiral-odd GPD $E^{u+d}_{T}$ at fixed $t = -0.5~\mathrm{GeV}^{2}$ with varying $\xi$ (upper panel), and at fixed $\xi = 0.25$ with varying $t$ (lower panel).}
\label{fig:5}
\end{figure}
The UV-divergent contributions in the gradient expansion to the mean-field GPDs affect only the ERBL regime, where $-\xi < x < \xi$. Therefore, we will focus solely on the non-forward limit of the GPDs. In addition to the leading contribution in the gradient expansion, there is an additional UV-divergent contribution at subleading order ($\alpha=1$). In the numerical calculations, we include all these contributions.

We first compute the mean-field GPDs and, using Eq.~\eqref{eq:largeNc_relation_nonsinglet}, extract the corresponding chiral-odd GPDs\footnote{In relating $Z_{\mathrm{mf},1}$ to the standard chiral-odd GPD $\bar{E}_{T}$, a careful matching is required within the accuracy of the present work. In the chiral expansion, the leading contribution to the GPD $H^{u-d}_{T}$ arises from UV-finite terms~\cite{inprep, Ghim:2025gqo}. Thus, the UV-divergent contribution appears only in the combination $\bar{E}^{u-d}_{T} \equiv 2\tilde{H}^{u-d}_{T} + E^{u-d}_{T}$. Therefore, the result for the mean-field GPD $Z_{\mathrm{mf},1}$ shoud be interpreted as
\begin{align}
Z_{\mathrm{mf},1}&= H^{u-d}_{T}+ 2\tilde{H}^{u-d}_{T} + E^{u-d}_{T} \nonumber \\[.5ex]
&=\bar{E}^{u-d}_{T} + \mathcal{O}(\text{UV-finite}).
\end{align}
On the other hand, when considering both the UV-finite and UV-divergent contributions in the analysis, the contribution from $H^{u-d}_{T}$ must be taken into account.}. In Figs.~\ref{fig:3}, \ref{fig:4}, and \ref{fig:5}, we present the numerical results for the chiral-odd GPDs $H_{T}$, $\bar{E}_{T}$, and $E_{T}$, respectively. All of these GPDs exhibit a convex, peak-like shape with a pronounced maximum at $x = 0$, which becomes sharper as $\xi$ decreases. The crossover from the ERBL to the DGLAP regime is smooth and continuous. The shape of this transition is mainly governed by the momentum-dependent dynamical quark mass at large quark virtuality, as given in Eq.~\eqref{eq:zmff} (see Refs.~\cite{Petrov:1998kf, Penttinen:1999th} for details).

In accordance with the $N_{c}$ scaling behavior of the chiral-odd GPDs, the typical magnitudes obtained in the numerical estimation follow the hierarchy:
\begin{align}
H^{u+d}_{T} \ll \bar{E}^{u-d}_{T} \ll E^{u+d}_{T}.
\label{eq:hier}
\end{align}
This hierarchy can also be understood in terms of multipole order. The magnitudes of $H^{u+d}_{T}$, $\bar{E}^{u-d}_{T}$, and $E^{u+d}_{T}$ are primarily determined by the monopole $Z_{\mathrm{mf},0}$, dipole $Z_{\mathrm{mf},1}$, and quadrupole $Z_{\mathrm{mf},2}$ mean-field GPDs, respectively. This implies that the strength of each GPD can be predicted from its underlying multipole structure.

Lastly, it is important to note that we observe a vanishing $\tilde{E}_{T}$ GPD, even when including the subleading contribution in the chiral expansion with $\alpha = 1$. The result for $\tilde{E}_{T}$ can be extracted from the relation $\tilde{E}_{T} = \tilde{Z}_{\mathrm{mf},1} + \xi Z_{\mathrm{mf},2}$ by solving Eqs.~\eqref{relation_1tilde} and \eqref{relation_2}. Numerically, we find that the mean-field GPDs $\tilde{Z}_{\mathrm{mf},1}$ and $Z_{\mathrm{mf},2}$ satisfy the relation:
\begin{align}
\tilde{Z}_{\mathrm{mf},1} = -\xi Z_{\mathrm{mf},2},
\end{align}
which leads to a vanishing $\tilde{E}_{T}$ GPD at the level of UV-divergent accuracy. To obtain a non-zero $\tilde{E}_{T}$, one must include contributions beyond the UV-divergent order.

\begin{figure}
\includegraphics[scale=0.30]{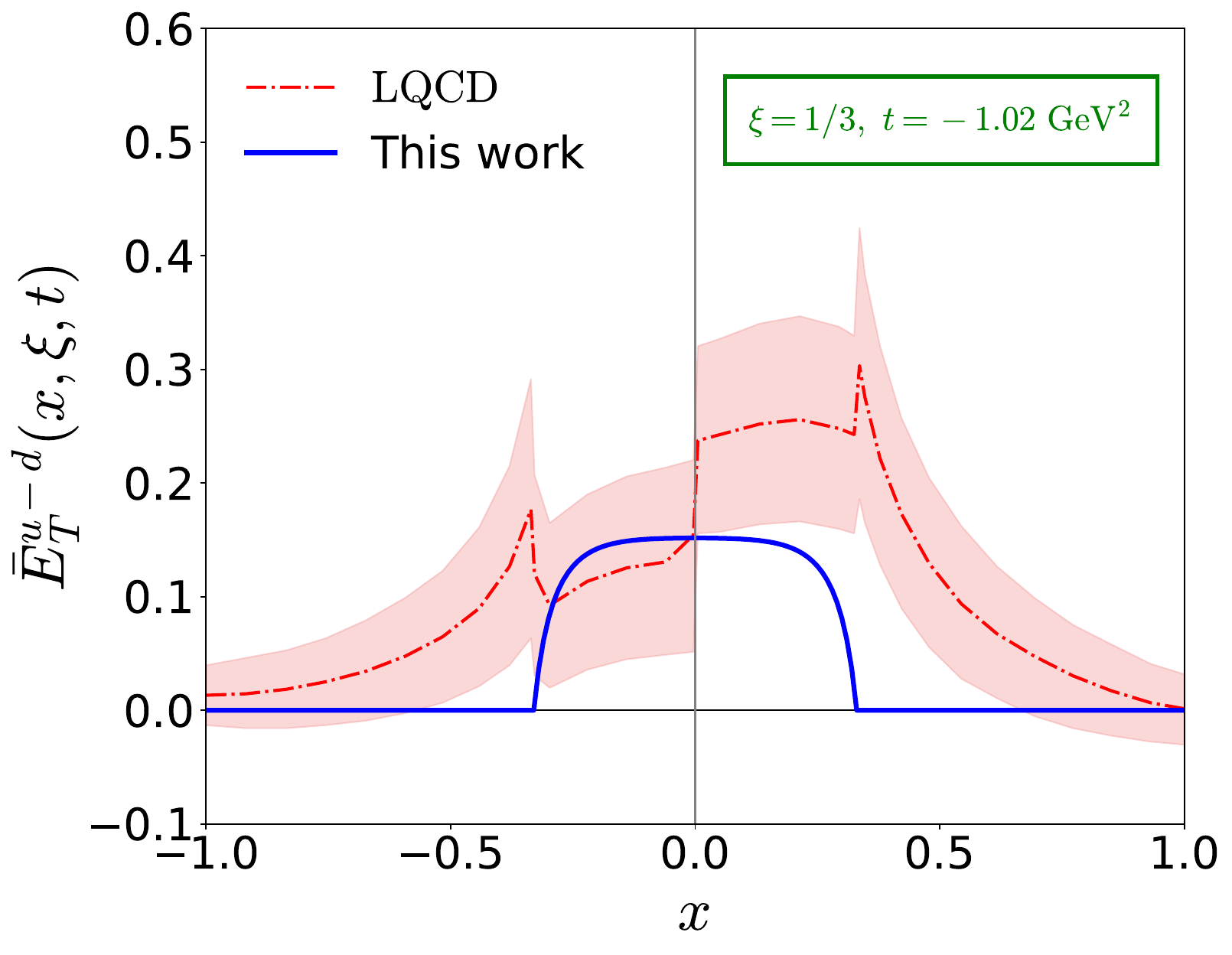} 
\caption{The result for the non-forward chiral-odd GPD $\bar{E}^{u-d}_{T}$ at $\xi = 1/3$ and $t = -1.02~\mathrm{GeV}^{2}$ is shown in comparison with the lattice QCD result. Our result is depicted as a solid (blue) curve, while the lattice QCD result from Ref.~\cite{Alexandrou:2021bbo} is shown as a dot-dashed (red) curve.}
\label{fig:6}
\end{figure}

Recently, non-forward chiral-odd GPDs have been computed exclusively in lattice QCD~\cite{Alexandrou:2021bbo, Bhattacharya:2025yba}, providing insights into their behavior in the ERBL regime. In particular, Ref.~\cite{Alexandrou:2021bbo} has calculated the ``light-cone” chiral-odd GPDs with non-zero skewness using quasi-GPDs through a perturbative matching procedure. Due to the challenges associated with disconnected diagrams, the study focuses on the flavor non-singlet component of the chiral-odd GPDs. Among these, the non-vanishing contribution to $\bar{E}^{u-d}_{T}$ appears at NLO in the $1/N_{c}$ expansion. We compare our result with the lattice QCD prediction in Fig.~\ref{fig:6}. In the negative-$x$ region, we observe good agreement with the lattice data, given the accuracy of the present work. In the positive-$x$ region, the slight discrepancy may be attributed to the absence of the discrete-level contribution (corresponding to the connected diagram in Fig.~\ref{fig:dia}), which requires a more complete analysis involving the full sum over single-particle wave functions.

\section{Summary and conclusions \label{sec:sac}}

In this work, we conducted a comprehensive study of chiral-odd GPDs within the mean-field framework, focusing on the first non-vanishing contributions that appear at next-to-leading order (NLO) in the $1/N_{c}$ expansion. This extends the leading-order (LO) results~\cite{Kim:2024ibz} and provides a complete set of predictions for both flavor-singlet and flavor-non-singlet chiral-odd GPDs.

In the mean-field framework, we performed zero-mode quantization (accounting for rotation and translation) and included the first-order rotational correction, thereby enabling the study of NLO contributions in the $1/N_{c}$ expansion. At  NLO, we evaluated the baryon matrix element of the chiral-odd QCD operator and derived the corresponding spin-flavor structures. Projecting these structures onto proton states and performing a multipole expansion, we obtained the associated multipole structure, where the mean-field/multipole GPDs encode the relevant dynamical information. We also determined their $N_{c}$ scaling behavior and, based on this, established the relation between the mean-field GPDs and the standard chiral-odd GPDs in the $1/N_{c}$ expansion.

The expressions for the mean-field GPDs were formulated in terms of matrix elements in the single-particle representation. By exploiting the symmetries of this representation—namely, parity and $G_{5}$ symmetry—within the mean-field framework, we computed the moments of the mean-field GPDs. Furthermore, leveraging the direct connection between the mean-field and standard chiral-odd GPDs, we demonstrated the polynomiality property and derived the associated sum rules for the chiral-odd GPDs by analyzing their mean-field counterparts. These results laid the groundwork for the subsequent numerical evaluation of the chiral-odd GPDs.

Using the gradient expansion and taking the nonvanishing UV-divergent contribution to the matrix element, we estimated the Dirac sea contribution to the chiral-odd GPDs. These contributions appear exclusively in the ERBL regime and are found to be smooth and continuous at the crossover points $x = \pm\xi$. Moreover, all of the chiral-odd GPDs exhibit a convex, peak-like shape with a pronounced maximum at $x = 0$, which becomes sharper as $\xi$ decreases. We numerically confirmed the large-$N_{c}$ hierarchy~\eqref{eq:hier} among the chiral-odd GPDs. In particular, we observed that $\tilde{E}^{u+d}_{T}$ does not receive a UV-divergent contribution, indicating the need for further investigation of its UV-finite part. Finally, a comparison with available lattice QCD data shows good agreement with our results.

Our study was limited to the UV-divergent contributions to the chiral-odd GPDs obtained via the gradient expansion, which affect only the non-forward GPDs. To achieve more accurate and complete predictions, it is necessary to sum over all single-particle wave functions, including contributions from discrete levels. Alternatively, one could compute the moments of the chiral-odd GPDs and use them to reconstruct the full distributions.

Finally, the spin-flavor operators generate not only the monopole operator but also the quadrupole spin operator~\eqref{eq:quad}, which provides additional dynamical information relevant for the $\Delta \to \Delta$ and $N \to \Delta$ transition matrix elements~\cite{Kim:2023yhp, Kim:2023xvw}.  This will be important for analyzing exclusive scattering amplitudes involving higher-spin particles~\cite{Semenov-Tian-Shansky:2023bsy, Kroll:2022roq}.

\section*{Acknowledgments}
JYK is indebted to C. Weiss for invaluable discussions, particularly for illuminating aspects of the effective field theory on the chiral-odd GPDs. JYK also wishes to express gratitude to N.Y. Ghim, H.Y. Won, H.D. Son, and H.-Ch. Kim for valuable discussions about the tensor form factors. This material is based upon work supported by the U.S.~Department of Energy, Office of Science, Office of Nuclear Physics under contract DE-AC05-06OR23177.

\appendix

\section{Tensor form factors \label{app:tff}}
In this section, we discuss the tensor form factors in the mean-field picture.
By considering the NLO matrix element~\eqref{eq:3corr} of the ``local" chiral-odd operator~\eqref{eq:eff} between proton states, we derive the following multipole structure:
\begin{subequations}
\label{eq:meanfield_FFs}
\begin{align}
& \mathcal{M}^{u-d}_{\mathrm{FFs}}[i \sigma^{0i}]=2 M_{N} \, \bm{1} \, Y^{i}_{1} \,
\frac{\sqrt{-t}}{2M_{N}} N_{\textrm{mf}, 1}, 
\\[1ex]
& \mathcal{M}^{u+d}_{\mathrm{FFs}}[i \sigma^{ij}]=2 M_{N} \bigg{[} i \epsilon^{ijm}\, \sigma^{m} \, Y_{0} \, 
N_{\textrm{mf}, 0}  
\cr
 &-i \epsilon^{ijl} \sigma^{m} \, Y^{lm}_{2} \, \frac{t}{4M^{2}_{N}} N_{\textrm{mf}, 2} \bigg{]},
\end{align}
\end{subequations}
in the 3D Breit frame, where $N_{\mathrm{mf}} \equiv N_{\mathrm{mf}}(t)$ denotes the mean-field (multipole) form factors appearing as rotational corrections. The explicit expressions for these are given by:
\begin{subequations}
\label{eq:local}
\begin{align}
&N_{\mathrm{mf},0} = - \frac{N_c}{6I} \sum_{\substack{n,\mathrm{non}  \\ j,\mathrm{occ} }} \frac{1}{E_n - E_j} \langle n | \tau^{m} | j \rangle \cr
&\times \langle j | \gamma_{0} \Sigma^{m} \,  j_{0}(|\bm{\hat{X}}| \sqrt{-t} ) | n \rangle, \\[1ex]
&N_{\mathrm{mf},1} = - \frac{iM_{N}N_c}{3I\sqrt{-t}} \sum_{\substack{n,\mathrm{non}  \\ j,\mathrm{occ} }} \frac{1}{E_n - E_j}  \langle n | \tau^{m} | j \rangle \cr
&\times  \langle j | \gamma^{k} \tau^{m}  \frac{\bm{\hat{X}}^{k}}{|\bm{\hat{X}}|}  j_{1}(|\bm{\hat{X}}| \sqrt{-t} )  | n \rangle, \\
&N_{\mathrm{mf},2} = - \frac{3M^{2}_{N}N_c}{It} \sum_{\substack{n,\mathrm{non}  \\ j,\mathrm{occ} }} \frac{1}{E_n - E_j}  \langle n | \tau^{l} | j \rangle \cr
&\times \langle j | \gamma_{0} \Sigma^{m} Y^{lm}_{2}\left(\Omega_{\bm{\hat{X}}}\right) j_{2}(|\bm{\hat{X}}| \sqrt{-t} ) \,   | n \rangle.
\end{align}
\end{subequations}
By performing the 3D multipole expansion of the covariant matrix element in Eq.~\eqref{eq:General_ME_tensor} and comparing the result with Eq.~\eqref{eq:local}, we obtain the following relations between the mean-field form factors and the tensor form factors:
\begin{subequations}
\label{eq:local_1}
\begin{align}
 H^{u+d}_{T} + \frac{t}{6M^{2}_{N}}E^{u+d}_{T}&=N_{\mathrm{mf},0}, \\
 H^{u-d}_{T}+E^{u-d}_{T}+2\tilde{H}^{u-d}_{T}&=N_{\mathrm{mf},1}, \\[1ex]
 E^{u+d}_{T}&=N_{\mathrm{mf},2}.
\end{align}
\end{subequations}
These combinations serve as the basis for proving polynomiality and sum rules, as well as for implementing numerical estimations.

\section{Polynomiality \label{app:poly}}

Using the results~\eqref{eq:moment} for the $m$-th moments of the mean-field GPDs in the single-particle representation~\eqref{eq:moment_1}, we have established the polynomiality of the monopole mean-field GPD in the large-$N_c$ limit. In this appendix, we extend the analysis to the higher-multipole chiral-odd GPDs and demonstrate their polynomiality properties. For simplicity, the proof is carried out in the limit $t \to 0$ with $\xi \neq 0$.

\subsection{$\xi$-even dipole GPD}
The single-particle matrix element $A_{1}$ for the $m$-th moment of the dipole GPD $Z_{\mathrm{mf},1}$ is given by
\begin{subequations}
\label{eq:even_dipole_1}
\begin{align}
&A^{ki}_1(\xi,t) = \frac{4}{3} \frac{\Delta^{a}}{|\bm{\Delta}_{\perp}|^{2}} \langle n| \tau^{b} | j \rangle  \cr
&\times \langle j | \tau^{b} (1+\gamma^{0} \gamma^{3}) \gamma^{a} (\hat{p}^{3})^{k-i}  e^{i\bm{\Delta} \cdot \hat{\bm{X}}} (\hat{p}^{3})^{i} | n \rangle, \\
&B^{ki}_1(\xi,t) = 4 \frac{\Delta^{a}}{|\bm{\Delta}_{\perp}|^{2}}  \cr
&\times \langle n | (1+\gamma^{0} \gamma^{3}) \gamma^{a} (\hat{p}^{3})^{k-i}  e^{i\bm{\Delta} \cdot \hat{\bm{X}}} (\hat{p}^{3})^{i} | n \rangle.
\end{align}
\end{subequations}
Considering the $G_{5}$ symmetry~\eqref{eq:Gparity}, Eq.~\eqref{eq:even_dipole_1} becomes
\begin{subequations}
\label{eq:even_dipole_2}
\begin{align}
&A^{ki}_1(\xi,t) = \frac{4}{3} \frac{\Delta^{a}}{|\bm{\Delta}_{\perp}|^{2}} \langle n| \tau^{b} | j \rangle  \cr
&\times \langle j | \tau^{b} (\gamma^{0} \gamma^{3})^{k} \gamma^{a} (\hat{p}^{3})^{k-i}  e^{i\bm{\Delta} \cdot \hat{\bm{X}}} (\hat{p}^{3})^{i} | n \rangle, \\
&B^{ki}_1(\xi,t) = 4 \frac{\Delta^{a}}{|\bm{\Delta}_{\perp}|^{2}}  \cr
&\times \langle n | (\gamma^{0} \gamma^{3})^{k} \gamma^{a} (\hat{p}^{3})^{k-i}  e^{i\bm{\Delta} \cdot \hat{\bm{X}}} (\hat{p}^{3})^{i} | n \rangle.
\end{align}
\end{subequations}
The operator $e^{i\bm{\Delta} \cdot \hat{\bm{X}}}$ in Eq.~\eqref{eq:even_dipole_2} is now expanded in terms of partial waves. In the limit $t \to 0$, the combination $\Delta^{a} e^{i\bm{\Delta} \cdot \hat{\bm{X}}}$ takes the form [cf.~\eqref{eq:pwe_fw}]
\begin{align}
&\Delta^{a} e^{i \bm{\Delta}\cdot \hat{\bm{X}}} \cr
&\overset{t\to0}{=} - i \nabla^{a} \left[\sum^{\infty}_{l=2} \frac{(-2i M_{N} \xi |\hat{\bm{X}}|)^{l}}{l!} P_{l} (\cos{\hat{\theta}}) \right],
\label{eq:even_dipole_3}
\end{align}
where the lowest value of $l$ that contributes nontrivially to Eq.~\eqref{eq:even_dipole_2} is $l = 2$. Using the relation~\eqref{eq:hierarchy2}, the squared transverse momentum transfer $|\bm{\Delta}_{\perp}|^{2}$ in the denominator of Eq.~\eqref{eq:even_dipole_2} becomes
\begin{align}
|\bm{\Delta}_{\perp}|^{2} \overset{t\to 0}{=} -(2\xi M_{N})^{2}.
\label{eq:even_dipole_4}
\end{align}
Substituting Eqs.~\eqref{eq:even_dipole_3} and \eqref{eq:even_dipole_4} into Eq.~\eqref{eq:even_dipole_2}, we obtain
\begin{subequations}
\label{eq:even_dipole_5}
\begin{align}
A^{ki}_1(\xi,0) &= -\frac{4}{3} i \sum^{\infty}_{l=2} \frac{(-2i M_{N} \xi)^{l-2}}{l!}  \langle n| \tau^{b} | j \rangle  \cr
&\times \langle j | \tau^{b} (\gamma^{0} \gamma^{3})^{k}  (\bm{\gamma}_{\perp} \cdot \bm{\nabla}_{\perp}) (\hat{p}^{3})^{k-i}   \cr
&\times|\hat{\bm{X}}|^{l} P_{l} (\cos{\hat{\theta}} ) (\hat{p}^{3})^{i} | n \rangle, \\[1ex]
B^{ki}_1(\xi,0) &= -4 i \sum^{\infty}_{l=2} \frac{(-2i M_{N} \xi)^{l-2}}{l!}   \cr
&\times \langle n |  (\gamma^{0} \gamma^{3})^{k}  (\bm{\gamma}_{\perp} \cdot \bm{\nabla}_{\perp}) (\hat{p}^{3})^{k-i}   \cr
&\times|\hat{\bm{X}}|^{l} P_{l} (\cos{\hat{\theta}} ) (\hat{p}^{3})^{i} | n \rangle.
\end{align}
\end{subequations}
Next, taking into account the parity invariance~\eqref{eq:parity}, we find that the allowed partial waves in Eq.~\eqref{eq:even_dipole_5} are restricted to even values, $l = 2, 4, \ldots$. This leads to the modification of the summation in Eq.~\eqref{eq:even_dipole_5} as
\begin{align}
\sum^{\infty}_{l=2}[...] \to \sum^{\infty}_{l=2,4, \ldots}[...].
\label{eq:even_dipole_6}
\end{align}
The part of the single-particle operator in Eq.~\eqref{eq:even_dipole_5} can be rewritten as
\begin{subequations}
\begin{align}
 (\gamma^{0}\gamma^{3})^{k} (\bm{\gamma}_{\perp} \cdot \bm{\nabla}_{\perp}) &= (\bm{\gamma}_{\perp} \cdot \bm{\nabla}_{\perp}) && (\text{$k$ even}), \\[1ex]
&=i  \gamma^{0}  (\bm{\Sigma} \times \bm{\nabla})^{3}  && (\text{$k$ odd}).
\label{eq:even_dipole_7}
\end{align}
\end{subequations}
Note that $\tau^{b}| j\rangle \langle j |\tau^{b}$ is invariant under hedgehog rotations. Therefore, it effectively behaves as a rank-zero 3D irreducible tensor operator with respect to simultaneous isospin and space rotations (cf.~\cite{Ossmann:2004bp}). Utilizing this property, along with the grand spin selection rule~\eqref{eq:Wigner_Ekart_1} for a single sum, we determine the maximum value of $l$ for the summation in Eq.~\eqref{eq:even_dipole_5}, following the same logic used in the derivation of Eq.~\eqref{l_max_m_2_main}:
\begin{subequations}
\label{eq:even_dipole_8}
\begin{alignat}{2}
l^{A,B}_{\mathrm{max}}(k) &= k + 2 \hspace{2em} && (\text{$k$ even}),
\\[1ex]
&= k +1 && (\text{$k$ odd}).
\end{alignat}
\end{subequations}
In terms of $m$, the maximum value is given by $l^{A}_{\mathrm{max}}(k = m - 1)$.
Furthermore, it is convenient to change the summation variable in Eq.~\eqref{eq:even_dipole_5} from $l$ to $l - 2$, so that the actual powers of $\xi$ appear in the polynomial. The maximum value of the new variable is then given by
\begin{subequations}
\label{l_max_m_2}
\begin{alignat}{2}
l^{A}_{\mathrm{max}}(m) &= m - 1 \hspace{2em} && (\text{$m$ odd}),
\\[1ex]
&= m -2 && (\text{$m$ even}).
\end{alignat}
\end{subequations}
The results for $l^{B}_{\mathrm{max}}(m)$ can also be simply obtained by replacing  $k = m - 2$. Shifting the summation variable $l$ accordingly, Eq.~\eqref{eq:even_dipole_5} becomes
\begin{subequations}
\label{eq:even_dipole_9}
\begin{align}
&A^{ki}_1(\xi,0) = -\frac{4}{3} i \sum^{l^{A}_{\mathrm{max}}(m)}_{l=0,2,4...} \frac{(-2i M_{N} \xi)^{l}}{(l+2)!}   \cr
&\times \langle n| \tau^{b} | j \rangle  \langle j | \tau^{b} (\gamma^{0} \gamma^{3})^{k}  (\bm{\gamma}_{\perp} \cdot \bm{\nabla}_{\perp}) \nonumber \\[1ex]
&\times (\hat{p}^{3})^{k-i} |\hat{\bm{X}}|^{l+2} P_{l+2} (\cos{\hat{\theta}} ) (\hat{p}^{3})^{i} | n \rangle, \\[1ex]
&B^{ki}_1(\xi,0) = -4 i \sum^{l^{B}_{\mathrm{max}}(m)}_{l=0,2,4 ...}  \frac{(-2i M_{N} \xi)^{l}}{(l+2)!}   \cr
&\times \langle n |  (\gamma^{0} \gamma^{3})^{k}  (\bm{\gamma}_{\perp} \cdot \bm{\nabla}_{\perp})    \cr
&\times (\hat{p}^{3})^{k-i} |\hat{\bm{X}}|^{l+2} P_{l+2} (\cos{\hat{\theta}} ) (\hat{p}^{3})^{i} | n \rangle.
\end{align}
\end{subequations}
We find that the dipole GPD $Z_{\mathrm{mf},1}$ is a polynomial in even powers of $\xi$, with its highest power bounded as required by Eq.~\eqref{eq:polynomiality}. This prove the polynomiality of the $\xi$-even dipole GPD.

\subsection{$\xi$-odd dipole GPD}

The single-particle matrix elements $\tilde{A}_{1}$ and $\tilde{B}_{1}$ for the $m$-th moment of the dipole GPD $\tilde{Z}_{\mathrm{mf},1}$ are given by
\begin{subequations}
\label{eq:odd_dipole_1}
\begin{align}
&\tilde{A}^{ki}_1(\xi,t) = 4 i \epsilon^{3ab} \frac{\Delta^{a}}{|\bm{\Delta}_{\perp}|^{2}} \langle n| \tau^{b} | j \rangle  \cr
&\times \langle j | (1+\gamma^{0} \gamma^{3}) \gamma^{a} (\hat{p}^{3})^{k-i}  e^{i\bm{\Delta} \cdot \hat{\bm{X}}} (\hat{p}^{3})^{i} | n \rangle,  \\[1ex]
&\tilde{B}^{ki}_1(\xi,t) = 4 i \epsilon^{3ab}  \frac{\Delta^{a}}{|\bm{\Delta}_{\perp}|^{2}} \cr
&\times \langle n | \tau^{b} (1+\gamma^{0} \gamma^{3}) \gamma^{a} (\hat{p}^{3})^{k-i} e^{i\bm{\Delta} \cdot \hat{\bm{X}}}(\hat{p}^{3})^{i} | n \rangle, 
\end{align}
\end{subequations}
Applying the $G_{5}$ symmetry~\eqref{eq:Gparity}, Eq.~\eqref{eq:odd_dipole_1} simplifies to
\begin{subequations}
\label{eq:odd_dipole_2}
\begin{align}
&\tilde{A}^{ki}_1(\xi,t) = 4 i \epsilon^{3ab} \frac{\Delta^{a}}{|\bm{\Delta}_{\perp}|^{2}} \langle n| \tau^{b} | j \rangle  \cr
&\times \langle j | (\gamma^{0} \gamma^{3})^{k+1} \gamma^{a} (\hat{p}^{3})^{k-i}  e^{i\bm{\Delta} \cdot \hat{\bm{X}}} (\hat{p}^{3})^{i} | n \rangle,  \\[1ex]
&\tilde{B}^{ki}_1(\xi,t) = 4 i \epsilon^{3ab}  \frac{\Delta^{a}}{|\bm{\Delta}_{\perp}|^{2}} \cr
&\times \langle n | \tau^{b} (\gamma^{0} \gamma^{3})^{k+1} \gamma^{a} (\hat{p}^{3})^{k-i} e^{i\bm{\Delta} \cdot \hat{\bm{X}}}(\hat{p}^{3})^{i} | n \rangle.
\end{align}
\end{subequations}
We now perform the partial wave expansion of the operator $\Delta^{a} e^{i\bm{\Delta} \cdot \hat{\bm{X}}}$ as outlined in Eq.~\eqref{eq:even_dipole_3}. Substituting Eqs.~\eqref{eq:even_dipole_3} and \eqref{eq:even_dipole_4} into Eq.~\eqref{eq:odd_dipole_2}, we obtain
\begin{subequations}
\label{eq:odd_dipole_5}
\begin{align}
&\tilde{A}^{ki}_1(\xi,0) \cr
&=4 \sum^{\infty}_{l=2} \frac{(-2i M_{N} \xi)^{l-2}}{l!}    \langle n| \tau^{3} | j \rangle \langle j   | (\gamma^{0} \gamma^{3})^{k+1}  \cr
&\times   (\bm{\gamma} \times \bm{\nabla})^{3} (\hat{p}^{3})^{k-i}  |\hat{\bm{X}}|^{l} P_{l} (\cos{\hat{\theta}} ) (\hat{p}^{3})^{i} | n \rangle,  \\[1ex]
&\tilde{B}^{ki}_1(\xi,0) \cr
&=4 \sum^{\infty}_{l=2} \frac{(-2i M_{N} \xi)^{l-2}}{l!} \langle n | \tau^{3} (\gamma^{0} \gamma^{3})^{k+1}  \cr
&\times    (\bm{\gamma} \times \bm{\nabla})^{3} (\hat{p}^{3})^{k-i} |\hat{\bm{X}}|^{l} P_{l} (\cos{\hat{\theta}} ) (\hat{p}^{3})^{i} | n \rangle.
\end{align}
\end{subequations}
Next, the parity transformation~\eqref{eq:parity} determines the allowed parity of the partial waves $l$. Under this transformation, only odd values of $l$ contribute to Eq.~\eqref{eq:odd_dipole_5}, modifying the summation as follows:
\begin{align}
\sum^{\infty}_{l=2}[\ldots] \to \sum^{\infty}_{l=3,5, \ldots}[\ldots].
\end{align}
Lastly, the maximal power of $l$ is constrained by the selection rules given in Eqs.~\eqref{eq:Wigner_Ekart_1} and \eqref{eq:Wigner_Ekart_2}. Using the relations in Eq.~\eqref{eq:gamma_i}, the grand spin selection rules, and the same reasoning applied in deriving Eq.~\eqref{l_max_m_2_main}, we determine the maximal value of $l$ to be
\begin{subequations}
\label{eq:odd_dipole_6}
\begin{alignat}{2}
l^{A,B}_{\mathrm{max}}(k) &= k+3  \hspace{2em} && (\text{$k$ even}),
\\[1ex]
&= k +2 && (\text{$k$ odd}).
\end{alignat}
\end{subequations}
This result can also be expressed in terms of the variable $m$ (with $k = m - 1$ for $l_{\mathrm{max}}^{A}$). Furthermore, by shifting the summation variable as $l \to l - 3$, we obtain:
\begin{subequations}
\label{eq:odd_dipole_8}
\begin{alignat}{2}
l^{A}_{\mathrm{max}}(m) &= m-1  \hspace{2em} && (\text{$m$ odd}),
\\[1ex]
&= m -2 && (\text{$m$ even}).
\end{alignat}
\end{subequations}
Similarly, the result for $l^{B}_{\mathrm{max}}$ follows by setting $k = m - 2$ in Eq.~\eqref{eq:odd_dipole_6}. Shifting the summation variable $l$ accordingly, Eq.~\eqref{eq:odd_dipole_5} becomes
\begin{subequations}
\label{eq:odd_dipole_9}
\begin{align}
&\tilde{A}^{ki}_1(\xi,0) \cr
&=4\sum^{l^{A}_{\mathrm{max}}(m)}_{l=0,2,4...} \frac{(-2i M_{N} \xi)^{l+1}}{(l+3)!}  \langle n| \tau^{3} | j \rangle \langle j | (\gamma^{0} \gamma^{3})^{k+1} \cr
&\times  (\bm{\gamma} \times \bm{\nabla})^{3} (\hat{p}^{3})^{k-i}  |\hat{\bm{X}}|^{l+3} P_{l+3} (\cos{\hat{\theta}} ) (\hat{p}^{3})^{i} | n \rangle, \\
&\tilde{B}^{ki}_1(\xi,0) \cr
&=4\sum^{l^{B}_{\mathrm{max}}(m)}_{l=0,2,4...} \frac{(-2i M_{N} \xi)^{l+1}}{(l+3)!}   \langle n | \tau^{3} (\gamma^{0} \gamma^{3})^{k+1} \cr
& \times (\bm{\gamma} \times \bm{\nabla})^{3} (\hat{p}^{3})^{k-i} |\hat{\bm{X}}|^{l+3} P_{l+3} (\cos{\hat{\theta}} ) (\hat{p}^{3})^{i} | n \rangle.
\end{align}
\end{subequations}
The $m$-th moment of the dipole mean-field GPD $\tilde{Z}_{\mathrm{mf},1}$ consists of contributions from $\xi E_{T}$ and $\tilde{E}_{T}$. As shown in Eq.~\eqref{eq:odd_dipole_9}, the dipole mean-field GPD is clearly an odd function of $\xi$, with maximal powers reaching $m$, as generally required by polynomiality from $\xi E_{T}$ and $\tilde{E}_{T}$. Thus, we have demonstrated the polynomiality of the $\xi$-odd dipole GPD $\tilde{Z}_{\mathrm{mf},1}$. Notably, the first moment of $\tilde{Z}_{\mathrm{mf},1}$ is nonzero due to the presence of the $\xi E_{T}$ term.

\subsection{Quadrupole GPD}

The single-particle matrix elements ${A}_{2}$ and ${B}_{2}$ for the $m$-th moment of the quadrupole GPD ${Z}_{\mathrm{mf},2}$ are given by
\begin{subequations}
\label{eq:quadrupole_1}
\begin{align}
&A^{ki}_{2}(\xi,t) = \frac{16 i \epsilon^{3ab}}{|\bm{\Delta}_{\perp}|^{4}} (\Delta^{b}_{\perp}\Delta^{c}_{\perp} - \frac{1}{2}\delta^{bc}|\bm{\Delta}_{\perp}|^{2})  \cr
&\hspace{-0.1cm}\times  \langle n| \tau^{c} | j \rangle \langle j | (1+\gamma^{0} \gamma^{3}) \gamma^{a} (\hat{p}^{3})^{k-i}  e^{i\bm{\Delta} \cdot \hat{\bm{X}}} (\hat{p}^{3})^{i} | n \rangle,  \\[1ex]
&B^{ki}_{2}(\xi,t) =  \frac{16 i \epsilon^{3ab} }{|\bm{\Delta}_{\perp}|^{4}} (\Delta^{b}_{\perp}\Delta^{c}_{\perp} - \frac{1}{2}\delta^{bc}|\bm{\Delta}_{\perp}|^{2}) \cr
&\times \langle n | \tau^{c} (1+\gamma^{0} \gamma^{3}) \gamma^{a} (\hat{p}^{3})^{k-i} e^{i\bm{\Delta} \cdot \hat{\bm{X}}}(\hat{p}^{3})^{i} | n \rangle, 
\end{align}
\end{subequations}
Applying the $G_{5}$ symmetry~\eqref{eq:Gparity}, Eq.~\eqref{eq:quadrupole_1} simplifies to
\begin{subequations}
\label{eq:quadrupole_2}
\begin{align}
&A^{ki}_{2}(\xi,t) =   \frac{16 i \epsilon^{3ab}}{|\bm{\Delta}_{\perp}|^{4}} (\Delta^{b}\Delta^{c} - \frac{1}{2}\delta^{bc}|\bm{\Delta}_{\perp}|^{2}) \cr
&\times \langle n| \tau^{c} | j \rangle \langle j | (\gamma^{0} \gamma^{3})^{k+1} \gamma^{a} (\hat{p}^{3})^{k-i}  e^{i\bm{\Delta} \cdot \hat{\bm{X}}} (\hat{p}^{3})^{i} | n \rangle,  \\[1ex]
&B^{ki}_{2}(\xi,t) =    \frac{16 i \epsilon^{3ab}}{|\bm{\Delta}_{\perp}|^{4}} (\Delta^{b}\Delta^{c} - \frac{1}{2}\delta^{bc}|\bm{\Delta}_{\perp}|^{2})\cr
&\times \langle n | \tau^{c} (\gamma^{0} \gamma^{3})^{k+1} \gamma^{a} (\hat{p}^{3})^{k-i} e^{i\bm{\Delta} \cdot \hat{\bm{X}}}(\hat{p}^{3})^{i} | n \rangle.
\end{align}
\end{subequations}
In the limit $t \to 0$, the projection of the partial waves $e^{i\bm{\Delta} \cdot \hat{\bm{X}}}$ onto the quadrupole structure is given by [cf.~\eqref{eq:pwe_fw}]
\begin{align}
&(\Delta^{b}\Delta^{c} - \frac{1}{2}\delta^{bc}|\bm{\Delta}_{\perp}|^{2})e^{i \bm{\Delta} \cdot \hat{\bm{X}}} \cr
&\overset{t\to 0}{=} -(\delta^{bp}\delta^{cq }- \frac{1}{2} \delta^{pq} \delta^{bc} )\cr
&\times \nabla^{p}_{\perp}\nabla^{q}_{\perp}  \left[\sum^{\infty}_{l=4} \frac{(-2i M_{N} \xi |\hat{\bm{X}}|)^{l}}{l!} P_{l} (\cos{\hat{\theta}}) \right], 
\label{eq:quadrupole_3_a}
\end{align}
where the minimal value of $l$ for a non-vanishing contribution to Eq.~\eqref{eq:quadrupole_2} is $l=4$. Inserting Eqs.~\eqref{eq:quadrupole_3_a} and \eqref{eq:even_dipole_4} into Eq.~\eqref{eq:quadrupole_2}, we obtain
\begin{subequations}
\label{eq:quadrupole_3}
\begin{align}
&A^{ki}_{2}(\xi,0)\cr
&= -16\sum^{\infty}_{l=4}  \frac{(-2i M_{N} \xi )^{l-4}}{l!}    (\delta^{bp}\delta^{cq }- \frac{1}{2} \delta^{pq} \delta^{bc} )\cr
&\times  i \epsilon^{3ab}  \langle n| \tau^{c} | j \rangle \langle j | (\gamma^{0} \gamma^{3})^{k+1} \gamma^{a}  \nonumber \\[1ex]
&\times (\hat{p}^{3})^{k-i}  \nabla^{p}_{\perp}\nabla^{q}_{\perp} |\hat{\bm{X}}|^{l}  P_{l} (\cos{\hat{\theta}}) (\hat{p}^{3})^{i} | n \rangle,  \\[1ex]
&B^{ki}_{2}(\xi,0) \cr
&= -16\sum^{\infty}_{l=4} \frac{(-2i M_{N} \xi )^{l-4}}{l!}  (\delta^{bp}\delta^{cq }- \frac{1}{2} \delta^{pq} \delta^{bc} )  \cr
&\times  i \epsilon^{3ab} \langle n | \tau^{c} (\gamma^{0} \gamma^{3})^{k+1} \gamma^{a}  \nonumber \\[1ex]
&\times (\hat{p}^{3})^{k-i} \nabla^{p}_{\perp}\nabla^{q}_{\perp} |\hat{\bm{X}}|^{l}  P_{l} (\cos{\hat{\theta}}) (\hat{p}^{3})^{i} | n \rangle, 
\end{align}
\end{subequations}
Next, applying the parity transformation discussed in Eq.~\eqref{eq:parity} to Eq.~\eqref{eq:quadrupole_3}, we find that only even powers of $l$ survive:
\begin{align}
\sum^{\infty}_{l=4}[\ldots] \to \sum^{\infty}_{l=4,6, \ldots}[\ldots].
\end{align}
To determine the upper bound of $l$, we examine the maximal rank of the single-particle operator. To see this, we first express the 2D irreducible tensor operator in terms of its 3D counterpart. For $A_{2}$ with even $k$, we have a irreducible rank-3 tensor:
\begin{align}
&i\epsilon^{3ab}\gamma^{0}\gamma^{3}\gamma^{a}\left(\nabla^{b}_{\perp}\nabla^{c}_{\perp} - \frac{1}{2}\delta^{bc} \bm{\nabla}^{2}_{\perp}\right) \cr
&= \gamma^{0}\Sigma^{b}_{\perp} (\nabla_{\perp}^{b}\nabla_{\perp}^{c}-\frac{1}{3} \delta^{bc} \bm{\nabla}^{2}) \cr
&+\frac{1}{2}\gamma^{0}\Sigma^{b}_{\perp} (\nabla^{3}\nabla^{3}-\frac{1}{3} \bm{\nabla}^{2})\delta^{bc}.
\label{eq:quadrupole_4}
\end{align}
Similarly, for odd $k$, we also obtain a irreducible rank-3 tensor:
\begin{align}
&\gamma^{a}\left(\nabla^{b}_{\perp}\nabla^{c}_{\perp} - \frac{1}{2}\delta^{bc}_{\perp} \bm{\nabla}^{2}_{\perp}\right) \cr
&= \gamma^{a} (\nabla_{\perp}^{b}\nabla_{\perp}^{c}-\frac{1}{3} \delta^{bc}_{\perp} \bm{\nabla}^{2}) \cr
&+ \frac{1}{2}\gamma^{a} (\nabla^{3}\nabla^{3}-\frac{1}{3} \bm{\nabla}^{2})\delta^{bc}_{\perp}.
\label{eq:quadrupole_5}
\end{align}
By coupling either Eq.~\eqref{eq:quadrupole_4} or Eq.~\eqref{eq:quadrupole_5} with the momentum/displacement operators in Eq.~\eqref{eq:quadrupole_3}, we determine the maximal rank of the single-particle operator. Following the reasoning used in the derivation of Eq.~\eqref{l_max_m_2_main}, we obtain the maximal value of $l^{A}_{\mathrm{max}}$:
\begin{align}
\label{eq:quadrupole_6}
l^{A}_{\mathrm{max}}(k) &= k+4,
\end{align}
which holds for all $k$. The corresponding results for $l^{B}_{\mathrm{max}}(k)$ can be derived in a similar way by considering Eqs.~\eqref{eq:quadrupole_4} and \eqref{eq:quadrupole_5} along with the selection rule for the single sum:
\begin{subequations}
\label{eq:quadrupole_7}
\begin{alignat}{2}
l^{B}_{\mathrm{max}}(k) &= k+4  \hspace{2em} && (\text{$k$ even}),
\\[1ex]
&= k +3 && (\text{$k$ odd}).
\end{alignat}
\end{subequations}
Rewriting Eqs.~\eqref{eq:quadrupole_6} and \eqref{eq:quadrupole_7} in terms of $m$ and shifting the summation variable $l \to l-4$, we obtain:
\begin{alignat}{2}
l^{A}_{\mathrm{max}}(m) &= m-1  \hspace{2em} && (\text{$k$ even/odd}). 
\label{eq:quadrupole_7_a}
\end{alignat}
The result for $l^{B}_{\mathrm{max}}(m)$ can be obtained analogously to Eq.~\eqref{eq:quadrupole_7_a}. Under the change of summation variable, Eq.~\eqref{eq:quadrupole_3} finally becomes:
\begin{subequations}
\label{eq:quadrupole_8}
\begin{align}
&A^{ki}_{2}(\xi,0) = -16\sum^{l^{A}_{\mathrm{max}}(m)}_{l=0,2,4...}  \frac{(-2i M_{N} \xi )^{l}}{(l+4)!}  i \epsilon^{3ab}\cr
&\times  (\delta^{bp}\delta^{cq }- \frac{1}{2} \delta^{pq} \delta^{bc} )   \langle n| \tau^{c} | j \rangle \langle j | (\gamma^{0} \gamma^{3})^{k+1} \gamma^{a}  \nonumber \\[1ex]
&\times (\hat{p}^{3})^{k-i}  \nabla^{p}_{\perp}\nabla^{q}_{\perp} |\hat{\bm{X}}|^{l+4}  P_{l+4} (\cos{\hat{\theta}}) (\hat{p}^{3})^{i} | n \rangle,  \\
&B^{ki}_{2}(\xi,0) = -16\sum^{l^{B}_{\mathrm{max}}(m)}_{l=0,2,4...} \frac{(-2i M_{N} \xi )^{l}}{(l+4)!}   i \epsilon^{3ab} \cr
&\times  (\delta^{bp}\delta^{cq }- \frac{1}{2} \delta^{pq} \delta^{bc} ) \langle n | \tau^{c} (\gamma^{0} \gamma^{3})^{k+1} \gamma^{a}  \nonumber \\[1ex]
&\times (\hat{p}^{3})^{k-i} \nabla^{p}_{\perp}\nabla^{q}_{\perp} |\hat{\bm{X}}|^{l+4}  P_{l+4} (\cos{\hat{\theta}}) (\hat{p}^{3})^{i} | n \rangle.
\end{align}
\end{subequations}
The $m$-th moment of the quadrupole mean-field GPD $Z_{\mathrm{mf},2}$ coincides with those of $E_{T}$ in the large-$N_{c}$ limit. As shown in Eq.~\eqref{eq:quadrupole_8}, the result is manifestly an even function of $\xi$, and the maximal power of $\xi$ reaches $m-1$, consistent with the polynomiality property of $E_{T}$. This confirms the polynomiality of $Z_{\mathrm{mf},2}$.

\section{Sum rules \label{app:sr}}

In this appendix, using the results from the polynomiality proof of the higher-multipole GPDs and taking their first moments ($m = 1$), we demonstrate that the mean-field GPDs are connected to the tensor form factors, in accordance with the standard sum rule for chiral-odd GPDs. As in Appendix~\ref{app:poly}, we verify the sum rules in the limit $t = 0$ with $\xi \neq 0$.

\subsection{$\xi$-even dipole GPD}
By taking the first moment $m = 1$ of the $\xi$-even dipole GPD in Eq.~\eqref{eq:moment}, we obtain
\begin{align}
&\int dx \, Z_{\mathrm{mf},1}(x, \xi, 0) \cr
&= -\frac{M_{N} N_{c}}{4I} \sum_{\substack{n,\mathrm{non} \\ j,\mathrm{occ}}} \frac{1}{E_{n}-E_{j}} A^{00}_{1}(\xi,0),
\label{eq:sr_ed1}
\end{align}
where the single-particle matrix element $A^{00}_{1}(\xi,0)$ is given by
\begin{align}
&A^{00}_{1}(\xi,0) = \frac{2}{3} i \langle n | \bm{\tau} | j \rangle \cdot \langle j | \bm{\tau} (\bm{\gamma_{\perp}} \cdot \bm{\hat{X}_{\perp}})| n \rangle,
\label{eq:sr_ed2}
\end{align}
where $\bm{\hat{X}}_{\perp}=(\hat{X}_{1},\hat{X}_{2})$. In the derivation of Eq.~\eqref{eq:sr_ed2}, we used the relation
\begin{align}
\nabla^{a}_{\perp} \left[ |\bm{\hat{X}}|^{2}P_{2}(\cos\hat{\theta}) \right] = -\hat{X}^{a}_{\perp}.
\end{align}
Owing to the minimal generalization of spherical symmetry induced by the hedgehog symmetry, along with the grand spin selection rule~\eqref{eq:Wigner_Ekart_2} for the double sum, the matrix element~\eqref{eq:sr_ed2} in the 2D representation can be related to a rotationally invariant form:
\begin{align}
 &\langle n | \bm{\tau} | j \rangle \cdot \langle j | \bm{\tau} (\bm{\gamma_{\perp}} \cdot \bm{\hat{X}_{\perp}})  | n \rangle  \cr
&= \frac{2}{3} \langle n | \bm{\tau} | j \rangle \cdot \langle j | \bm{\tau} (\bm{\gamma } \cdot \bm{\hat{X} })  | n \rangle .
\end{align}
Using the relation above, we can express the first moment of the $\xi$-even dipole GPD~\eqref{eq:sr_ed1} in terms of the mean-field tensor form factors~\eqref{eq:local}, derived from the local operator. We then obtain
\begin{align}
&\int dx \, Z_{\mathrm{mf},1}(x, \xi, 0) = N_{\mathrm{mf},1}(0). 
\end{align}
Using the relation~\eqref{eq:local_1} between the mean-field and tensor form factors, we arrive at
\begin{align}
&\int dx \, Z_{\mathrm{mf},1}(x, \xi, 0) \cr
&=H^{u-d}_{T}(0)+E^{u-d}_{T}(0)+2\tilde{H}^{u-d}_{T}(0).
\end{align}
This result coincides with the first moment of the mean-field GPD defined in Eq.~\eqref{relation_1}. Thus, we have proven the sum rule for the $\xi$-even dipole mean-field GPD.

\subsection{$\xi$-odd dipole GPD}
By taking the first moment $m = 1$ of the $\xi$-odd dipole GPD in Eq.~\eqref{eq:moment}, we obtain
\begin{align}
&\int dx \, \tilde{Z}_{\mathrm{mf},1}(x, \xi, 0) \cr
&= -\frac{M_{N}N_{c}}{4I} \sum_{\substack{n,\mathrm{non} \\ j,\mathrm{occ}}} \frac{1}{E_{n}-E_{j}} \tilde{A}^{00}_{1}(\xi,0),
\label{eq:sr_od1}
\end{align}
where the single-particle matrix element $\tilde{A}^{00}_{1}(\xi,0)$ is given by
\begin{align}
&\tilde{A}^{00}_{1}(\xi,0) \cr
&= 4 M_{N} \xi \langle n | \tau^{3} | j \rangle \cdot \langle j | \gamma^{0} \Sigma^{a}_{\perp} |\bm{\hat{X}}|Y^{a3}_{2} (\Omega_{\hat{\bm{X}}}) | n \rangle 
\label{eq:sr_od2}
\end{align}
In deriving Eq.~\eqref{eq:sr_od2}, we used the relations for the Dirac matrices~\eqref{eq:gamma_i} and
\begin{align}
\nabla^{a} \left[ |\bm{\hat{X}}|^{3}P_{3}(\cos\hat{\theta}) \right] = -3|\bm{\hat{X}}|^{2}Y^{a3}_{2}(\Omega_{\hat{\bm{X}}}).
\end{align}
Using the grand spin selection rule~\eqref{eq:Wigner_Ekart_2} for the double sum, we rewrite Eq.~\eqref{eq:sr_od2} in a 3D rotationally invariant form:
\begin{align}
 &\langle n | \tau^{3} | j \rangle  \langle j |  \gamma^{0} \Sigma^{b}_{\perp} |\bm{\hat{X}}|^{2} Y^{b3}_{2}(\Omega_{\hat{\bm{X}}}) | n \rangle  \cr
&=  \frac{1}{5}\langle n | \tau^{a} | j \rangle  \langle j |  \gamma^{0} \Sigma^{b} |\bm{\hat{X}}|^{2} Y^{ab}_{2}(\Omega_{\hat{\bm{X}}}) | n \rangle.
\end{align}
Using this relation, we express the first moment of the $\xi$-odd mean-field GPD~\eqref{eq:sr_od1} in terms of the mean-field tensor form factors~\eqref{eq:local}. After that, taking into account the relation~\eqref{eq:local_1}, we find
\begin{align}
&\int dx \, \tilde{Z}_{\mathrm{mf},1}(x, \xi, 0) =-\xi E^{u+d}_{T}(0).
\end{align}
This result coincides with the first moment of the mean-field GPD defined in Eq.~\eqref{relation_1tilde}, where the first moment of $\tilde{E}_{T}$ is zero. We have thus proven the sum rule for the $\xi$-odd dipole mean-field GPD.

\subsection{Quadrupole GPD}
By taking the first moment $m = 1$ of the $\xi$-odd dipole GPD in Eq.~\eqref{eq:moment}, we derive
\begin{align}
&\int dx \, Z_{\mathrm{mf},2}(x, \xi, 0) \cr
&= -\frac{M^{2}_{N}N_{c}}{4I} \sum_{\substack{n,\mathrm{non} \\ j,\mathrm{occ}}} \frac{1}{E_{n}-E_{j}} A^{00}_{2}(\xi,0),
\label{eq:sr_q1}
\end{align}
where the single-particle matrix element $A^{00}_{2}(\xi,t)$ is found to be
\begin{align}
&A^{00}_{2}(\xi,0) = -\frac{2}{3}  (\delta^{bp}\delta^{cq }- \frac{1}{2} \delta^{pq} \delta^{bc} ) \cr
&\times \langle n | {\tau}^{c} | j \rangle \langle j | \gamma^{0}\Sigma^{a}_{\perp} \nabla^{p}_{\perp}\nabla^{q}_{\perp} |\hat{\bm{X}}|^{4} P_{4}(\cos{\hat{\theta}}) | n \rangle.
\label{eq:sr_q2}
\end{align}
In deriving Eq.~\eqref{eq:sr_q2}, we used the relations for the Dirac matrices~\eqref{eq:gamma_i}. Using the identity
\begin{align}
&\left(\delta^{bp}\delta^{cq }- \frac{1}{2} \delta^{pq} \delta^{bc} \right) \nabla^{p}_{\perp} \nabla^{q}_{\perp} \left[ |\bm{\hat{X}}|^{4}P_{4}(\cos\hat{\theta}) \right]  \cr
&= 3 \left( \hat{X}^{b}_{\perp}\hat{X}_{\perp}^{c} - \frac{1}{2} \delta^{bc} |\hat{\bm{X}}_{\perp}|^{2}\right).
\end{align}
and reducing the 2D irreducible tensor to its 3D counterpart [cf.~\eqref{eq:3dto2d}], we arrive at
\begin{align}
&A^{00}_{2}(\xi,0) \cr
&= -2 \langle n | \tau^{a}_{\perp} | j \rangle  \langle j |  \gamma^{0} \Sigma^{b}_{\perp} |\bm{\hat{X}}|^{2} Y^{ab}_{2}(\Omega_{\hat{\bm{X}}}) | n \rangle  \cr
 &-\langle n | \tau^{a}_{\perp} | j \rangle  \langle j |  \gamma^{0} \Sigma^{a}_{\perp} |\bm{\hat{X}}|^{2} Y^{33}_{2}(\Omega_{\hat{\bm{X}}}) | n \rangle .
\label{eq:sr_q3}
\end{align}
Considering the grand spin selection rule~\eqref{eq:Wigner_Ekart_2}, we obtain the following identities:
\begin{subequations}
\begin{align}
 &\langle n | \tau^{a}_{\perp} | j \rangle  \langle j |  \gamma^{0} \Sigma^{b}_{\perp} |\bm{\hat{X}}|^{2} Y^{ab}_{2}(\Omega_{\hat{\bm{X}}}) | n \rangle  \cr
&= \frac{7}{15} \langle n | \tau^{a} | j \rangle  \langle j |  \gamma^{0} \Sigma^{b} |\bm{\hat{X}}|^{2} Y^{ab}_{2}(\Omega_{\hat{\bm{X}}}) | n \rangle, \\[1ex]
 &\langle n | \tau^{a}_{\perp} | j \rangle  \langle j |  \gamma^{0} \Sigma^{b}_{\perp} |\bm{\hat{X}}|^{2} Y^{33}_{2}(\Omega_{\hat{\bm{X}}}) | n \rangle  \cr
&= -\frac{2}{15} \langle n | \tau^{a} | j \rangle  \langle j |  \gamma^{0} \Sigma^{b} |\bm{\hat{X}}|^{2} Y^{ab}_{2}(\Omega_{\hat{\bm{X}}}) | n \rangle.
\end{align}
\end{subequations}
These relations allow us to restore the 3D rotationally invariant expression in Eq.~\eqref{eq:sr_q3}. Thus, the first moment of the quadrupole GPD~\eqref{eq:sr_q1} is related to the mean-field form factors~\eqref{eq:local}. Employing the relation~\eqref{eq:local_1}, we finally obtain
\begin{align}
&\int dx \, Z_{\mathrm{mf},2}(x, \xi, 0) = E^{u+d}_{T}(0).
\end{align}
This relation coincides with the first moment of the mean-field GPD defined in Eq.~\eqref{relation_2}. We have thus proven the sum rule for the quadrupole mean-field GPD.

\bibliography{chiral_odd_gpds}
\end{document}